\documentclass[aps,prl,superscriptaddress,amsfonts,amsmath,amssymb,showpacs,floatfix,reprint,longbibliography]{revtex4-2}
\usepackage{txfonts}
\usepackage{graphicx}
\usepackage{xcolor}
\usepackage[version=4]{mhchem}
\usepackage{multirow}
\usepackage{siunitx}

\begin{document}

\title{Dynamic Jahn-Teller effect in the strong spin-orbit coupling regime}

\author{Ivica {\v Z}ivkovi{\' c}}
\email{ivica.zivkovic@epfl.ch}
\affiliation{Laboratory for Quantum Magnetism, Institute of Physics, \'Ecole Polytechnique F\'ed\'erale de Lausanne, CH-1015 Lausanne, Switzerland}

\author{Jian-Rui Soh}
\affiliation{Laboratory for Quantum Magnetism, Institute of Physics, \'Ecole Polytechnique F\'ed\'erale de Lausanne, CH-1015 Lausanne, Switzerland}

\author{Oleg Malanyuk}
\affiliation{Laboratory for Quantum Magnetism, Institute of Physics, \'Ecole Polytechnique F\'ed\'erale de Lausanne, CH-1015 Lausanne, Switzerland}

\author{Ravi Yadav}
\affiliation{Institute of Physics, \'Ecole Polytechnique F\'ed\'erale de Lausanne, CH-1015 Lausanne, Switzerland}

\author{Federico Pisani}
\affiliation{Laboratory for Quantum Magnetism, Institute of Physics, \'Ecole Polytechnique F\'ed\'erale de Lausanne, CH-1015 Lausanne, Switzerland}

\author{Aria M. Tehrani}
\affiliation{Department of Materials, ETH Zurich, CH-8093 Zurich, Switzerland}

\author{Davor Tolj}
\affiliation{Laboratory for Quantum Magnetism, Institute of Physics, \'Ecole Polytechnique F\'ed\'erale de Lausanne, CH-1015 Lausanne, Switzerland}

\author{Jana Pasztorova}
\affiliation{Laboratory for Quantum Magnetism, Institute of Physics, \'Ecole Polytechnique F\'ed\'erale de Lausanne, CH-1015 Lausanne, Switzerland}

\author{Daigorou Hirai}
\affiliation{Department of Applied Physics, Nagoya University, Nagoya 464-8603, Japan}

\author{Yuan Wei}
\affiliation{Paul Scherrer Institute, Villigen PSI, Switzerland}

\author{Wenliang Zhang}
\affiliation{Paul Scherrer Institute, Villigen PSI, Switzerland}

\author{Carlos Galdino}
\affiliation{Paul Scherrer Institute, Villigen PSI, Switzerland}

\author{Tianlun Yu}
\affiliation{Paul Scherrer Institute, Villigen PSI, Switzerland}

\author{Kenji Ishii}
\affiliation{Synchrotron Radiation Research Center, National Institutes for Quantum Science and Technology, Sayo, Hyogo 679-5148, Japan}

\author{Albin Demuer}
\affiliation{Universit\'e Grenoble Alpes, INSA Toulouse, Universit\'e Toulouse Paul Sabatier, CNRS, LNCMI, F-38000 Grenoble, France}

\author{Oleg V. Yazyev}
\affiliation{Chair of Computational Condensed Matter Physics, Institute of Physics, \'Ecole Polytechnique F\'ed\'erale de Lausanne, CH-1015 Lausanne, Switzerland}

\author{Thorsten Schmitt}
\affiliation{Paul Scherrer Institute, Villigen PSI, Switzerland}

\author{Henrik M. R{\o}nnow}
\affiliation{Laboratory for Quantum Magnetism, Institute of Physics, \'Ecole Polytechnique F\'ed\'erale de Lausanne, CH-1015 Lausanne, Switzerland}

\date{\today}

\begin{abstract}
Exotic quantum phases, arising from a complex interplay of charge, spin, lattice and orbital degrees of freedom, are of immense interest to a wide research community. A well-known example of such an entangled behavior is the Jahn-Teller effect, where the lifting of orbital degeneracy proceeds through lattice distortions, often accompanied by ordering of spins and metal-insulator transitions. Static distortions, including cooperative behavior, have been associated with colossal magneto-resistance, multiferroicity, high-$T_\mathrm{C}$ superconductivity and other correlated phenomena. Realizations of the dynamic Jahn-Teller effect, on the other hand, are scarce since the preservation of vibronic symmetries requires subtle tuning of the local environment. Here we demonstrate that a highly-symmetrical 5d$^1$ double perovskite Ba$_2$MgReO$_6$, comprising of a 3D array of isolated ReO$_6$ octahedra, fulfils these requirements, resulting in a unique case of a dynamic Jahn-Teller system with strong spin-orbit coupling. Thermodynamic and resonant inelastic x-ray scattering experiments undoubtedly show that the Jahn-Teller instability leads to a ground-state doublet, invoking a paradigm shift for this family of compounds. The restoration of vibronic degrees of freedom arises from a quantum-mechanical zero-point motion, as revealed by detailed quantum chemistry calculations. The dynamic state of ReO$_6$ octahedra persists down to the lowest temperatures, where a multipolar order sets in, allowing for investigations of the interplay between a dynamic JT effect and strongly correlated electron behavior.
\end{abstract}

\maketitle

The importance of the Jahn-Teller (JT) effect reaches across many areas of chemistry, physics and applied sciences~\cite{Bersuker2006}. Examples range from localized systems like photo-excited methane cations~\cite{Li2021,Ridente2023}, ultra-fast transients in molecular complexes~\cite{Barlow2022} and vacancy diffusion in graphene~\cite{Babar2018} towards more cooperative phenomena including high-temperature superconductivity in cuprates~\cite{Keller2008}, colossal magneto-resistance in manganites~\cite{Zheng2003}, metal-insulator transition in ruthenates~\cite{Vitalone2022} and antiferroelectricity in lacunar spinels~\cite{Geirhos2021}. The corner-stone of the JT effect is the existence of an electronic state with orbital degeneracy, leading to an overall decrease of the energy once the degeneracy is lifted. Typically, this occurs through ligand ($L$) displacement, which consequently affects both electronic and magnetic properties of materials. In strongly correlated systems, where ligands are shared between neighbouring metal ($M$) sites, the distortions at one site are correlated with distortions on neighbouring sites, leading to a cooperative JT effect~\cite{Khomskii2021}.


\begin{figure*}
\centering
\includegraphics[width=1.9\columnwidth]{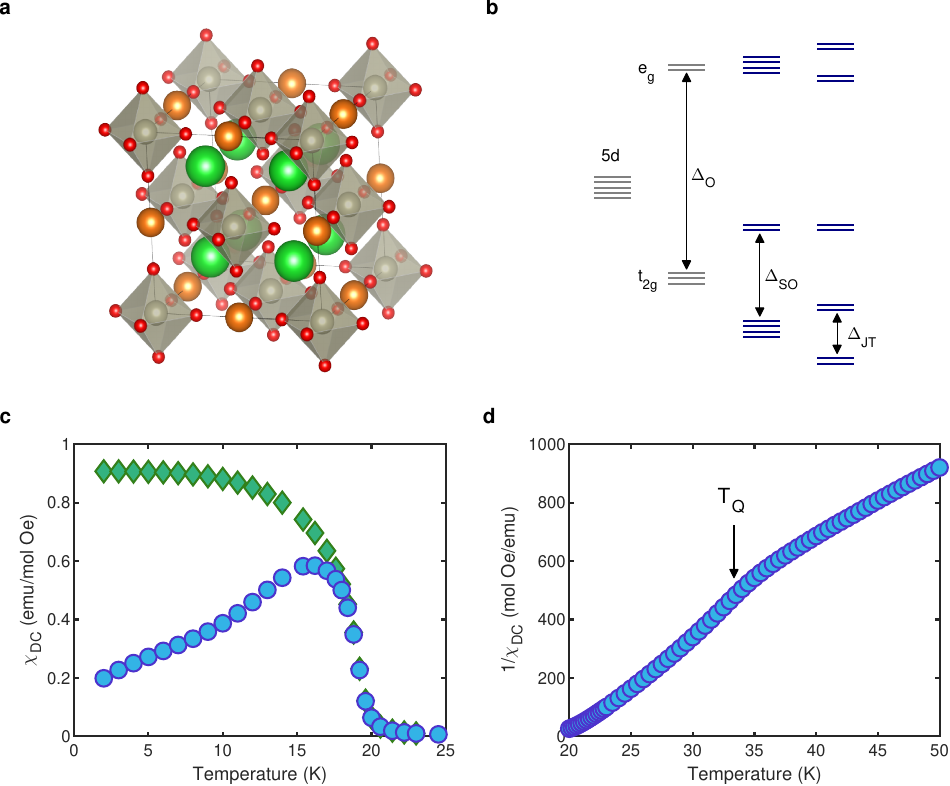}
\caption{\textbf{The crystal environment of \ce{Ba2MgReO6} and the multipolar order.} (a) The unit cell of \ce{Ba2MgReO6} with semi-transparent \ce{ReO6} octahedra. Mg$^{2+}$ and Ba$^{2+}$ ions are represented by orange and green circles, respectively. (b) Schematic representation of the splitting of Re 5$d$-levels under influence of crystal field ($\Delta_\mathrm{O}$), spin-orbit coupling ($\Delta_\mathrm{SO}$) and a JT effect ($\Delta_\mathrm{JT}$). Light grey lines represent orbitals, each allowing for two spin states. Dark blue lines represent SOC-entangled states. (c) $T$-dependence of dc magnetic susceptibility measured with $B = 0.1$\,T after zero-field-cooled (blue circles) and field-cooled (green diamonds) protocols. The transition temperature into long-range magnetic order occurs at $T_\mathrm{M} = 18$\,K and is marked by the splitting of two curves. (d) $T$-dependence of the inverse susceptibility with the quadrupolar order indicated as a change of the slope.}
\label{fig:INTRO}
\end{figure*}


In the case of isolated, highly-symmetric clusters the system can choose among equivalent directions of distortion, resulting in a potential energy surface (PES) with multiple valleys and energy barriers between them. In the simplest case of an $ML_{6}$ octahedron having a single electron or a hole in a doubly-degenerate orbital state (well known cases are Mn$^{3+}$ and Cu$^{2+}$ ions), the distortion can freeze along principal axes \textit{x}, \textit{y} and \textit{z}, producing a static JT effect. If the barriers are small enough so that the wave-function is distributed over all valleys equally, the vibronic symmetry is restored, resulting in a dynamic JT effect. Many aspects of solid-state physics have been theoretically investigated through the coupling to continuous symmetries in JT systems~\cite{Millis1996,Kayanuma2017,Ribeiro2018}, but the available materials are rather scarce. The inter-cluster interaction needs to be very small to avoid a cooperative JT effect, while other couplings, like exchange interactions, should remain appreciable. Examples of materials where a dynamic JT effect has been suggested to occur include an unconventional molecular superconductor \ce{Cs3C60}, where, together with Mott localization, it drives the material across a metal-to-insulator transition~\cite{Klupp2012,Iwahara2013,Zadik2015}. It has also been argued that the observed permittivity~\cite{Wieczorek2006} in a paraelectric phase of \ce{BaTiO3} stems from the resonating state between minima of the PES which under the influence of an electric field become nonequivalent and trap Ti ions along one direction~\cite{Bersuker2015}. Eventually, the dynamic JT effect is overcome by the interaction between the centers, leading to a ferroelectric phase transition.

These examples involve relatively light elements, including 3$d$ transition metal cations, where the influence of spin-orbit coupling (SOC) is considered negligible. Recently, a growing interest is devoted to strong SOC systems, encompassing topological insulators~\cite{Hasan2010}, Dirac and Weyl semimetals~\cite{Armitage2018}, and Kitaev spin liquids~\cite{Takagi2019}. The JT effect, on the other hand, has been seldom investigated in the context of strong SOC, with several examples devoted to Ir$^{4+}$ systems~\cite{Plotnikova2016,Liu2019}. More generally, a strong SOC qualitatively changes the ground-state wave-functions, and has been proposed to renormalize the associated distortion, or even completely remove it above a certain critical value~\cite{Streltsov2020}.


\begin{figure*}
\centering
\includegraphics[width=1.9\columnwidth]{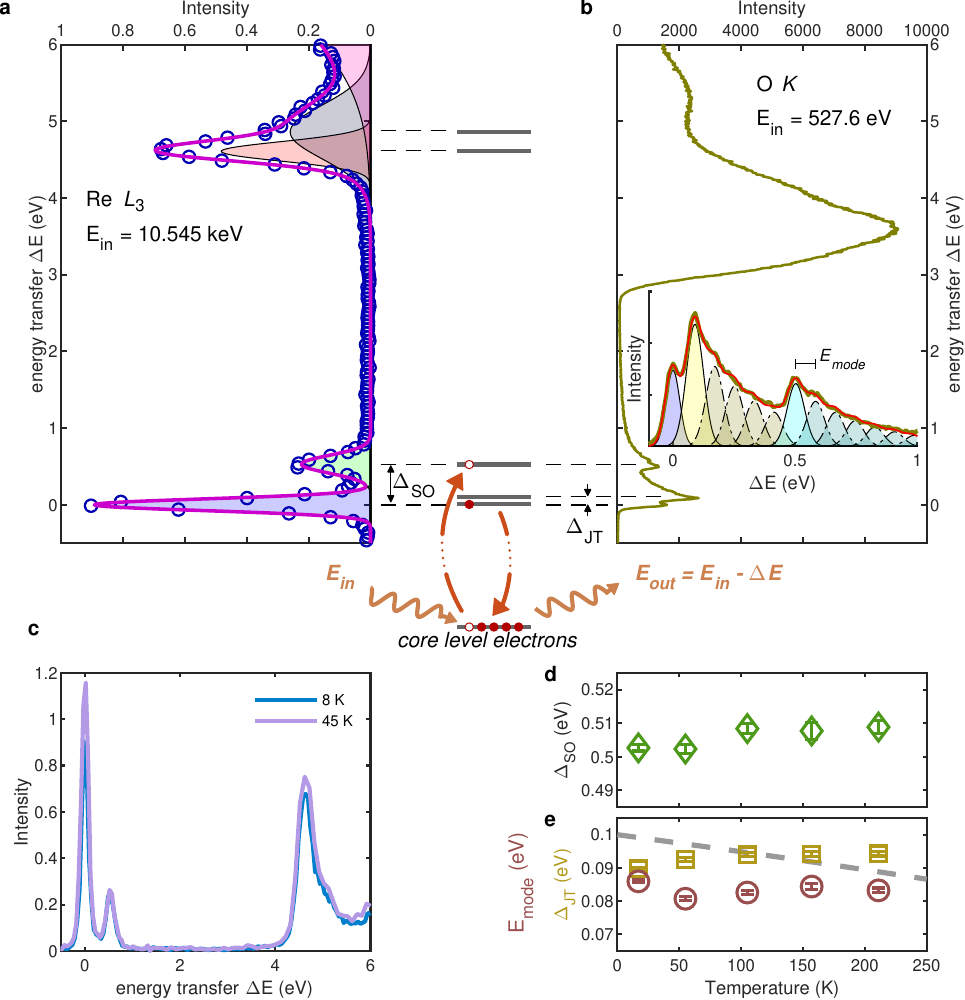}
\caption{\textbf{Resonant inelastic x-ray scattering on \ce{Ba2MgReO6}} (a) RIXS spectra at the Re \textit{L}$_3$ edge. The full line represents a sum of Gaussian contributions centered around the positions of localized 5$d$ Re levels. (b) RIXS spectra at the O \textit{K} edge. The excitations below 1\,eV correspond to localized $d-d$ processes while the broad features at higher energies are arising from fluorescence. The inset shows an expanded region below 1\,eV with highlighted contribution from individual modes. The RIXS scattering involves a virtual process of core-level electron excitation with a photon of energy $E_\mathrm{in}$ while the outgoing photon's energy $E_\mathrm{out}$ is reduced by the energy difference $\Delta E$, measuring the position of higher-lying $d$-levels. (c) Re RIXS spectra below and above $T_\mathrm{Q}$. (d) $T$-dependence of $\Delta_\mathrm{SO}$. (e) $T$-dependence of $\Delta_\mathrm{JT}$ and $E_{mode}$. The dashed line represents a hypothetic mean-field order parameter behavior $\Delta_\mathrm{JT} \sim \sqrt{1 - T/T_\mathrm{JT}}$ with $T_\mathrm{JT} = 1000$\,K.}
\label{fig:RIXS}
\end{figure*}


In this study, we present compelling evidence that \ce{Ba2MgReO6}, a member of the $5d^1$ double-perovskite family, displays a distinctive set of properties that make it exceptionally well-suited for the exploration of the role of the dynamic JT effect on strongly correlated electrons in the presence of a strong SOC. This is promoted by a high-symmetry environment of Re$^{6+}$ ions, covalently coupled to six oxygen ions, alternating with ionic Mg$^{2+}$ ions and forming a three-dimensional array of nearly-independent, JT-active \ce{ReO6} octahedra, Figure~\ref{fig:INTRO}a. Strong SOC mixes the $t_{2g}$ levels into a $j_\mathrm{eff} = 3/2$ quartet and a $j_\mathrm{eff} = 1/2$ doublet, separated by $\Delta_\mathrm{SO}$, as shown in Figure~\ref{fig:INTRO}b. The JT instability manifests itself as the splitting of the ground-state quartet into two doublets, with the gap $\Delta_\mathrm{JT}$ directly related to the amplitude of the distortion. The correlated nature of 5$d$ electrons is reflected in the formation of a multipolar order~\cite{Hirai2019,Hirai2020}, encompassing a charge quadrupole state below $T_\mathrm{Q} = 33$\,K and a magnetic dipole state below $T_\mathrm{M} = 18$\,K (Figure~\ref{fig:INTRO}c-d).

Using spectroscopic and thermodynamic evidence we show that the JT instability causes the doublet-doublet splitting in \ce{Ba2MgReO6} at temperatures $T >> T_\mathrm{M}, T_\mathrm{Q}$. We quantify the associated distortions through local cluster calculations and determine that the energy barriers between the local minima are smaller than the zero-point motion energy, establishing \ce{Ba2MgReO6} as a dynamic JT system. Additionally, through a careful analysis of specific heat results, we establish that the total entropy across \textit{both} transitions reaches only $R \mathrm{ln} 2$.


\begin{figure*}
\centering
\includegraphics[width=1.9\columnwidth]{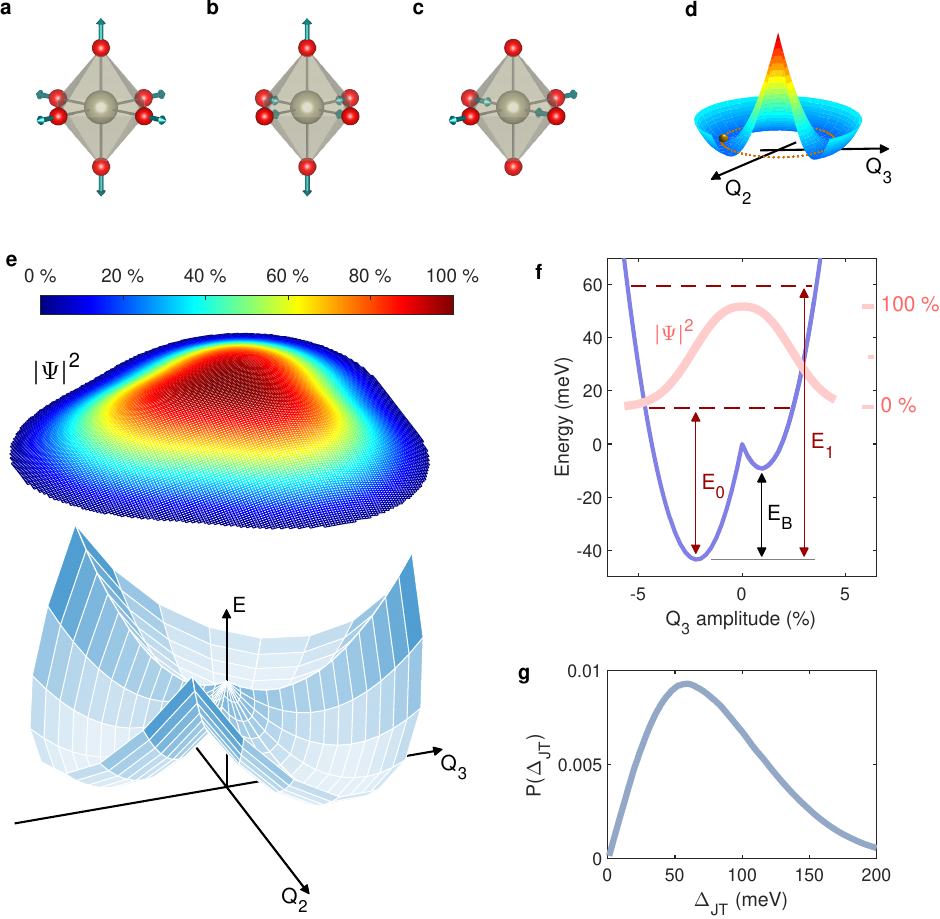}
\caption{\textbf{Vibrational modes and potential energy surface for a \ce{ReO6} cluster} Definition of displacement modes: (a) isotropic mode $\boldsymbol{Q}_{iso} = (1,1,1)$, (b)  $\boldsymbol{Q_3} = (-\frac{1}{2},-\frac{1}{2},1)$ and (c) $\boldsymbol{Q_2} = (-1,1,0)$. (d) A Mexican-hat potential with an infinite line degeneracy. (e) The PES obtained using quantum chemistry calculations. Superimposed on top is the numerically calculated probability density $|\Psi|^2$ of the ground-state wave function $\Psi (E = E_0)$ with a cut-off at 20\,\% of the maximum value. (f) A cross-section of the PES for $Q_2$ amplitude equal 0. (g) A probability distribution of $\Delta_\mathrm{JT}$ calculated from $P(\Delta_\mathrm{JT}) = \int \Delta_\mathrm{JT}(Q_i)|\Psi(Q_i)|^2 dQ_i$.}
\label{fig:QCC}
\end{figure*}


To probe the localized states of Re ions we employ resonant inelastic x-ray scattering (RIXS). In Figure~\ref{fig:RIXS}a we display the low temperature ($T = 8$\,K) energy dependence of the scattering intensity obtained at the Re \textit{L}$_3$ edge, displaying multiple peaks of varying widths, which we assign as follows (see also the supplementary material). The peak around zero energy is an elastic line, where no energy transfer occurs during the virtual excitation of core level electrons. Further peaks reflect virtual processes where core level electrons are excited to unoccupied states while the recombination of the core level hole occurs with electrons on occupied Re 5$d$ states. We find that the observed scattering intensity profile can be fully described by assuming that each peak is represented by a simple Gaussian contribution. The positions of maxima correspond to crystal field levels of Re 5$d$-orbitals and can be readily assigned following the scheme presented in Figure~\ref{fig:INTRO}b. The first peak at $\Delta E > 0$ represents the SOC-induced splitting between the ground state quartet and the higher-lying doublet state $\Delta_\mathrm{SO} \cong 0.5$\,eV which allows the estimate of the SOC constant $\lambda \sim 335$\,meV. Up to 4\,eV the line-shape is featureless, with two additional peaks present around $4.5 - 5$\,eV. A notable tail coming from an even higher lying peak can be observed, which can be assigned to virtual processes involving the recombination with electrons originating from lower-lying, predominantly oxygen \textit{p}-orbitals, located $\sim 6.3$\,eV below the elastic line. A recent report~\cite{Frontini2023} focusing on the comparison between \ce{Ba2MgReO6}, \ce{Sr2MgReO6} and \ce{Ca2MgReO6} found a slightly asymmetric line-shape around $\Delta_\mathrm{SO}$, ascribing it to the influence of phonon modes on spin-orbit excitations. Nevertheless, their value for $\Delta_\mathrm{SO}$ in \ce{Ba2MgReO6}, as well as the energies of higher-lying excitations, are in excellent agreement with our results.

The two peaks, situated at 4.7\,eV and 4.9\,eV, can be assigned to states belonging to $e_g$ orbitals, with four-fold degeneracy being split into two doublets, in accordance with the JT scenario. An equivalent split of the ground state quartet is expected but not directly observed due to an insufficient energy resolution ($\sim 100$\,meV) of the experimental setup at the Re $L_3$ edge. On the other hand, the results obtained at the O \textit{K} edge (Figure~\ref{fig:RIXS}b), with its $\sim 50$\,meV resolution, show an additional peak at low energies reflecting the doublet-doublet splitting $\Delta_\mathrm{JT} \cong 88$\,meV. A similar finding has been reported very recently~\cite{Agrestini2024} for the sister compound \ce{Ba2NaOsO6}, with the corresponding excitation at 95\,meV. It is interesting to note that both experiments at the O \textit{K} edge resulted in a similarly asymmetric line-shape around $\Delta_\mathrm{JT}$ and $\Delta_\mathrm{SO}$, with possible faint features on the high-energy side. The description of those features has recently been attempted in terms of coupling of phonon modes with spin-orbit excitations~\cite{Frontini2023}, although the asymmetry itself is still not fully understood. It is clear that several additional modes are necessary to describe such an asymmetric line-shape but if all the parameters are left independent, it quickly becomes unfeasible to perform the fitting procedure (see the supplementary material for more details). The approach we used is to model the high-energy tail of each $d-d$ excitation through a set of $n$ dependent modes, whose energies and amplitudes are determined by simple relations, $E_n^{\alpha} = E_{\alpha} + n \cdot E_{mode}$ and $A_n^{\alpha} = A_0^{\alpha} e^{-n \cdot B}$, with all the additional modes having the same width. Here $\alpha$ stands for a JT or SO $d-d$ excitation (indicated by light colors in the inset of Figure~\ref{fig:RIXS}b), while additional modes are indicated by darker colors. Such an approach significantly reduces the number of independent parameters and leads to a very satisfactory description of the observed line-shape within the relevant energy range up to 1\,eV.

The central result of RIXS experiments comes from the temperature evolution of the acquired spectra. In Figure~\ref{fig:RIXS}c we compare the Re $L_3$ RIXS spectra above and below $T_\mathrm{M}$ and $T_\mathrm{Q}$. Within the resolution there is no visible variation between the two line profiles. Specifically, the asymmetric shape at $4.5 - 5$\,eV, associated with the splitting of $e_g$ orbitals, remains clearly unaffected. A more direct and profound evidence comes from $\Delta_\mathrm{SO} (T)$ and $\Delta_\mathrm{JT} (T)$, extracted from O $K$ RIXS spectra and plotted in panels~\ref{fig:RIXS}d-e. Both are practically $T$-independent, with a small decrease of $\Delta_\mathrm{JT}$ below $T_\mathrm{Q}$.

It is not surprising that $\Delta_\mathrm{SO}$ shows very little temperature dependence. On the contrary, the persistence of $\Delta_\mathrm{JT}$ well above 200\,K goes against the prevailing idea that the splitting of the ground-state quartet is linked with the appearance of the quadrupolar order at $T_\mathrm{Q}$~\cite{Chen2010,Lu2017}. Moreover, we can infer that $\Delta_\mathrm{JT} > 0$ to much higher temperatures since any order-parameter-like decrease would be readily observed in our data, as indicated in Figure~\ref{fig:RIXS}e by a dashed line.

Now we turn our efforts to characterize the type of distortion that accompanies the observed splitting. The relevant modes of an $ML_{6}$ octahedron are sketched in Figure~\ref{fig:QCC}a-c. The magnitude of the isotropic mode mainly determines the octahedral splitting $\Delta_\mathrm{O}$ which for ~\ce{Ba2MgReO6} is large enough to disregard the $t_{2g}$$-$$e_g$ mixing. Two other modes are JT-active, with $Q_3$ representing a tetragonal distortion while $Q_2$ describes the orthorhombic mode. For an isolated octahedron, with a linear coupling and in the limit of very small distortions, the PES is characterized by an infinite-degeneracy line, the so-called 'Mexican hat' potential (Figure~\ref{fig:QCC}d). The octahedron remains dynamic, exploring all possible linear combinations in the $Q_2$$-$$Q_3$ parameter space, with $\Delta_\mathrm{JT} = const. > 0$. Increasing the amplitude of distortions leads to anharmonics effects and, together with the presence of surrounding charges, breaks the line degeneracy, resulting in a multi-valley PES. An opposing effect has been argued to arise from strong SOC~\cite{Streltsov2020}, driving the PES back towards a Mexican-hat potential but also suppressing the JT distortion, as well as reducing the energy gain due to the JT effect.


\begin{table*}
\caption{\textbf{Calculated energy levels for various structures.} JT-distorted octahedra are characterized by a non-zero value of $\Delta_\mathrm{JT}$ and two values for $\Delta_\mathrm{O}$. Columns refer to (from left to right): experimentally obtained ion positions at 40\,K; ion positions after a 1\% isotropic contraction of \ce{ReO6} octahedron; ion positions at the global minima, which includes the amplitude of the $Q_3$ mode of $-$1.5\%; experimental values obtained from Figure~\ref{fig:RIXS}; a fictitious tetragonal structure with 0.15\% elongation.}

\begin{tabular}{ | c || c | c | c | c || c | }
 \hline

 \multirow{2}{3em}{(meV)} & $Fm\overline{3}m$ & isotropic & global & \multirow{2}{3em}{\textbf{EXP}} & \multirow{2}{4em}{$I_4/mmm$} \\

 & @ 40\,K & contraction ($-$1\%) & minimum & & \\

 \hline
 \hline
 $\Delta_\mathrm{JT}$ & 0 & 0 & 78 & \textbf{88} & 1 \\  
 \hline
 $\Delta_\mathrm{SO}$ & 560 & 561 & 606 & \textbf{528} & 584 \\
 \hline
 \multirow{2}{2em}{$\Delta_\mathrm{O}$} & \multirow{2}{2em}{5490} & \multirow{2}{2em}{5573} & 5577 & \textbf{4608} & 5512 \\

 & & & 5962 & \textbf{4861} & 5523 \\
  \hline
\end{tabular}
\label{tbl:Elevels}
\end{table*}


For a more quantitative insight, we perform embedding multi-reference quantum chemistry calculations taking into account SOC (see Methods). We start from the experimentally established structure at 40\,K~\cite{Hirai2020} and calculate the PES of the \ce{ReO6} octahedron by parameterizing ligand displacements in terms of the above-described modes for a general distortion $\boldsymbol{\tau} = \alpha \boldsymbol{Q}_{iso} + \beta \boldsymbol{Q_3} + \gamma \boldsymbol{Q_2}$. In Figure~\ref{fig:QCC}e we show the calculated PES exhibiting three valleys, corresponding to tetragonal distortions along the three principal axes. One of the minima is given by $\alpha \approx -1$\%, $\beta \approx -1.5$\% and $\gamma = 0$, where $\beta < 0$ indicates that the degeneracy lifting of the ground-state quartet occurs through a \textit{contraction} of the Re$-$O$_{apical}$ distance. A similar result has been obtained for \ce{Ba2NaOsO6}~\cite{Agrestini2024} while calculations performed for a 4$d^1$ compound \ce{Ba2YMoO6} indicate the same sign and a similar magnitude of distortions~\cite{Xu2016}, thus establishing a clear tendency of $d^1$ octahedral systems towards a contraction, regardless of the strength of SOC. Importantly, both the sign and the magnitude of the calculated JT distortion are in stark contrast with the reported shift of oxygen ions below $T_\mathrm{Q}$~\cite{Hirai2020}. Detailed synchrotron x-ray diffraction analysis revealed~\cite{Hirai2020} the lowering of the global symmetry from a cubic $Fm\overline{3}m$ above $T_\mathrm{Q}$ to a tetragonal $P4_2/mnm$ below $T_\mathrm{Q}$, with a tetragonal \textit{elongation} amounting to $\sim 0.1$\% at low temperatures. This separation of scales undoubtedly disentangles the formation of the quadrupolar order at $T_\mathrm{Q}$ from the dominant, JT-induced splitting of the presumed ground-state quartet~\cite{Chen2010,Svoboda2021,Iwahara2023}.

The validity of the performed calculations can be assessed by comparing the calculated energy levels with experimental data as listed in Table~\ref{tbl:Elevels}. The calculations for the experimentally established structure at 40\,K and the isotropically contracted one contain undistorted octahedra, resulting in $\Delta_\mathrm{JT} = 0$ as well as four-fold degenerate $e_g$ states (see Figure~\ref{fig:INTRO}). The calculations for distortions characterizing the minimum of a valley lead to $\Delta_\mathrm{JT} = 78$\,meV and the splitting of $e_g$ states, a hall-mark of the JT effect. All the calculated observables are found to be in the range $10 - 20$\,\% from the experimental values.

The minimum of the PES does not directly reflect the amplitude of the distortion of an octahedron. We use the calculated PES to obtain the wave-function $\Psi$ of the quantum-mechanical oscillator and its ground-state energy $E_0$. As shown in Figure~\ref{fig:QCC}e, the wave function reflects the three-fold symmetry of the PES, extending towards each valley but still maintaining the absolute maximum at the center of the PES. More importantly, the energy $E_0 \approx 56$\,meV is found well above the barrier between valleys $E_B \approx 35$\,meV (Figure~\ref{fig:QCC}f), implying that the zero-point motion prevents the freezing of vibronic degrees of freedom, which leaves each octahedron in a dynamic state. The characteristic time is typically expressed in the form $(t_0)^{-1} = (E_1 - E_0)/h$, where $E_1 \approx 102$\,meV is the first excited state~\cite{Razavy2014}. With $t_0 \approx 9 \cdot 10^{-14}$\,s, only the scattering processes of x-rays and neutrons are fast enough to observe an instantaneous state of the system, all other experimental techniques obtain a dynamically averaged response.

We can calculate the probability distribution of $\Delta_\mathrm{JT}$ using $S(\Delta_\mathrm{JT}) = \int \Delta_\mathrm{JT}(Q_i)|\Psi(Q_i)|^2 dQ_i$, where $dQ_i$ designates an integration over the $Q_2 - Q_3$ space. The result presented in Figure~\ref{fig:QCC}g reveals an asymmetric shape with a maximum occurring around 60\,meV. We should emphasize that despite numerous approximations that underlie calculations, the results are directly comparable to experimentally obtained values. For a comparison, the last column in Table~\ref{tbl:Elevels} contains the results for a fictitious tetragonal $I_4/mmm$ structure, where a small 0.15\,\% tetragonal \textit{elongation} is implemented to reflect the observed cubic-to-tetragonal transition at $T_\mathrm{Q}$~\cite{Hirai2020}. The calculated $\Delta_{\mathrm{JT}}$ is two orders of magnitude too small, unambiguously assigning the observed doublet-doublet splitting to a JT instability.


\begin{figure*}
\centering
\includegraphics[width=1.9\columnwidth]{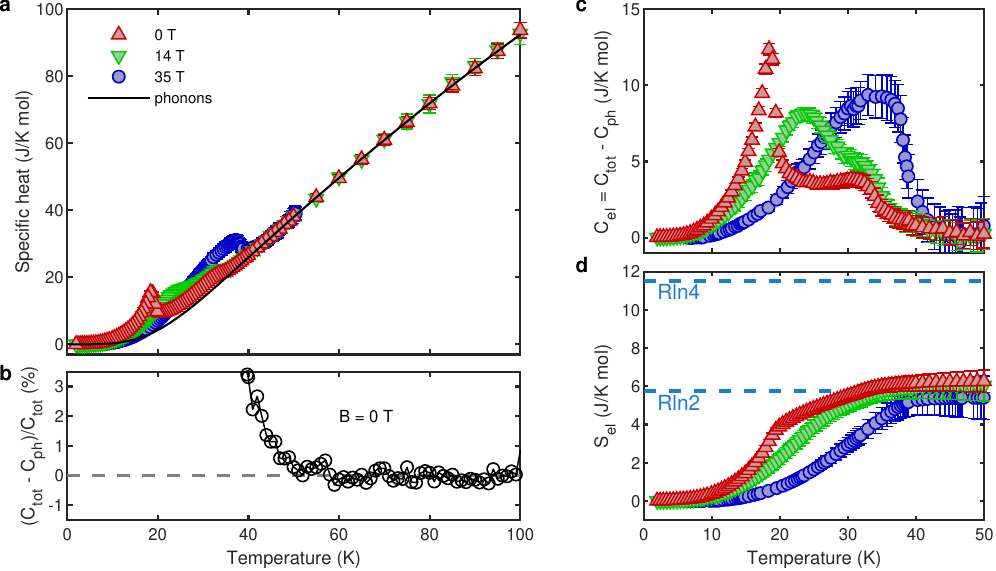}
\caption{\textbf{Specific heat measurements of \ce{Ba2MgReO6}.} (a) $T$-dependence of the total specific heat of \ce{Ba2MgReO6}. (b) The relative error of the electronic specific heat after the phonon background subtraction. (c) $T$-dependence of the electronic specific heat in various magnetic fields. (d) $T$-dependence of recovered entropy.}
\label{fig:SPECIFICHEAT}
\end{figure*}


These results can be naturally extended to other members of the double perovskite family where bonding with ligands alternates regularly between an ionic and a covalent type. Many similarities exist with \ce{Ba2NaOsO6}, which exhibits much smaller $T_\mathrm{M} = 7.5$\,K and $T_\mathrm{Q} = 9.5$\,K~\cite{Erickson2007,Lu2017,Willa2019}. Another example is \ce{Cs2TaCl6} where a tetragonal contraction of $< 1$\,\% has also been associated with the quadrupolar order~\cite{Ishikawa2019}, although a later single crystal study indicated a different value for $T_\mathrm{Q}$~\cite{Mansouri2023}. Several $d^2$ compounds seem to retain high local symmetry~\cite{Maharaj2020}, as well as a $d^3$ \ce{Ba2YOsO6} system~\cite{Kermarrec2015}, offering a variety of configurations where the interplay of a (dynamic) JT effect and strong SOC can be tested~\cite{Streltsov2020}.

The JT-splitting of the ground-state quartet has profound consequences for thermodynamic properties of 5$d^1$ systems. Already the first report on \ce{Ba2NaOsO6} indicated a 'missing entropy'~\cite{Erickson2007}, extracted from electronic specific heat ($C_\mathrm{el}$) measurements $S_\mathrm{el} = \int (C_\mathrm{el}/T)dT$ reaches only $\lesssim R \mathrm{ln}2$, significantly smaller than the expected $R \mathrm{ln}4$. Namely, the total recovered entropy is expected to reflect the ground-state degeneracy at high temperatures, released across two successive transitions $T_\mathrm{M}$ and $T_\mathrm{Q}$, each contributing one $R \mathrm{ln}2$~\cite{Chen2010,Svoboda2021,Iwahara2023}. Experimentally, the main uncertainty comes from the determination of a proper phonon background, which leads to a wide array of recovered values of $S_\mathrm{el}$~\cite{Marjerrison2016,Willa2019,Hirai2019,Ishikawa2021,Pasztorova2023}. We approach the problem of the phonon background determination by measuring specific heat $C_P$ in magnetic fields which, by affecting the electronic levels, causes a redistribution of measured $C_P$. Since the \textit{g}-factor is strongly renormalized due to the partial cancellation of spin and orbital moments~\cite{Chen2010}, it is necessary to apply large magnetic fields to see those effects around $T_\mathrm{Q}$ and above. In Figure~\ref{fig:SPECIFICHEAT}c we plot $C_P (T)$ of \ce{Ba2MgReO6} in magnetic fields up to 35\,T. For $B = 0$ a sharp peak is seen at $T_\mathrm{M}$ and a broad shoulder around $T_\mathrm{Q}$, in agreement with previously published results~\cite{Hirai2019,Pasztorova2023}. For $B = 35$\,T a broad feature is seen around $T_\mathrm{Q}$, with the tail merging with $C_P(B = 0\,\textrm{T})$ above $\sim 45$\,K. We take this as an estimate up to which temperature the electronic contribution is appreciable and consider only the phonon contribution above it. Instead of an analytical form, we aim to utilize the results of first-principle calculations which were recently used as an approximate description of the phonon specific heat in \ce{Ba2MgReO6}~\cite{Pasztorova2023}. We use a functional rescaling of the calculated results along both $T-$ and $C_P-$ axes to match the experimental data in the range 60 -- 100\,K (see the supplementary material for more details). The result of this procedure is presented in Figure~\ref{fig:SPECIFICHEAT}a as a black line, with the relative discrepancy remaining below 0.5\% across the relevant range (see Figure~\ref{fig:SPECIFICHEAT}b).

The electronic contribution $C_\mathrm{el}$ extracted in this manner is displayed in Figure~\ref{fig:SPECIFICHEAT}c. Both features at $T_\mathrm{M}$ and $T_\mathrm{Q}$ are more clearly seen, as well as the magnetic field dependence. Initially $T_\mathrm{M}$ becomes rounded and shifts to higher temperatures, while there is very little effect on $T_\mathrm{Q}$. With $B = 35$\,T, however, we see the two features practically merged, with a sharp high-temperature tail. Qualitatively similar behavior has been observed in CeB$_6$~\cite{Peysson1986,Amara2020} and assigned to field-induced octupolar moments on Ce ions~\cite{Shiina1997,Matsumara2009}. Similarly, the NMR study~\cite{Lu2017} revealed a shift to higher temperatures of the order-disorder boundary in Ba$_2$NaOsO$_6$.

The total electronic entropy $S_\mathrm{el}$ extracted from Figure~\ref{fig:SPECIFICHEAT}c is presented in Figure~\ref{fig:SPECIFICHEAT}d. Within an error, the recovered values for all magnetic fields saturate very close to $R \mathrm{ln}2$, with very little change above 40\,K. This result is rather robust against small changes of the exact functional form of the phonon background and therefore, together with spectroscopic evidence obtained by RIXS and supported by quantum chemistry calculations, unequivocally establishes that the low energy physics in \ce{Ba2MgReO6} is dominated by a ground state \textit{doublet}. A similar conclusion has been reached by a recent study on \ce{Ba2NaOsO6} where a somewhat larger doublet-doublet splitting has been found (95\,meV) based on O \textit{K} edge RIXS and x-ray magnetic circular dichroism~\cite{Agrestini2024}. We note that these two 5$d^1$ systems exhibit qualitatively very similar behavior in terms of multi-polar ordering tendencies, with the advantage that Re$^{+6}$ orders at sufficiently larger temperatures than Os$^{+7}$ to allow a better insight into the peculiar interplay between magnetic dipole and charge quadrupole orders.


\begin{figure*}
\centering
\includegraphics[width=1.9\columnwidth]{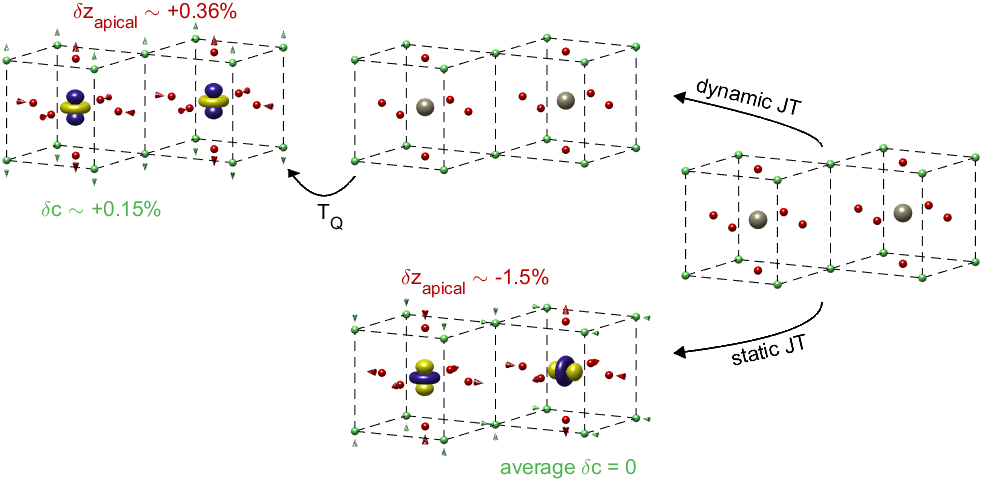}
\caption{\textbf{Comparison of dynamic and static JT effect on equilibrium ion positions at time-scales longer than $t_0$} At very high temperatures the octahedra are not distorted and the ground state is four-fold degenerate. In the dynamical JT scenario the degeneracy is lowered by a contraction of Re - O distances ($\sim -1.5$\,\%) on time-scales $t < t_0$ but the equilibrium positions of ions at $t > t_0$ do not change significantly, with the Re charge distribution remaining spherical. At much lower temperatures ($T < T_Q$) a quadrupolar order sets in, including the ferroic $\mathcal{T}_z$ (blue - increased charge density, yellow - decreased charge density), which is accompanied by a coherent shift of apical oxygen ions, leading to a small $c$-axis elongation ($\sim +0.15$\,\%)~\cite{Hirai2020}. In the static JT scenario each octahedron is locally distorted, with equal distribution of elongations along three principal axes. This implies a random 
distribution of $\mathcal{T}_z$ quadrupolar moments and a lack of coherent ordering at low temperatures. The local Ba cage is also distorted, leading to the accumulation of lattice strain.}
\label{fig:ISO}
\end{figure*}


The calculated ground-state energy suggests a dynamic type of distortion, where \ce{ReO6} octahedra fluctuate with $\sim 10$\,THz between elongations along three equivalent principal axes. The scattering of x-ray photons, which occurs at the time scale much faster than these oscillations, reflects a statistical average of all the distortions distributed across the sample at a given time and therefore cannot be used to distinguish it from a random distribution of frozen static distortion, which might still preserve the global cubic symmetry, despite the breaking of the local one. To the contrary, the static version of the JT effect can be reasonably argued against using several lines of reasoning (see Figure~\ref{fig:ISO} for details). First, there is a clear signature of order-parameter-like increase of intensity associated with charge quadrupolar order composed of two components, an antiferro $\mathcal{T}_x$ and a ferro $\mathcal{T}_z$~\cite{Hirai2020,Soh2023}. Concomitantly, it is observed that oxygen ions move away from their high-symmetry positions accordingly, reflecting a strong coupling between local distortions and $\mathcal{T}_z$ charge quadrupoles~\cite{Chen2010,Mansouri2021}. Those distortions are, however, 5 times smaller than calculated for the Jahn-Teller effect, and of an opposite sign. The frozen landscape of static JT distortions would inevitably lead to a glassy behavior of a charge quadrupolar and magnetic dipole orders, since the direction of magnetic dipoles below $T_Q$ is determined by the local orientation of $\mathcal{T}_z$ quadrupolar moments~\cite{Mansouri2021}.

The second argument for the dynamic JT state is related to the local symmetry of Re ions. As has been demonstrated in $^{23}$Na NMR experiments on \ce{Ba2NaOsO6}, a single resonance line at high temperatures splits into three lines below $T_Q$, indicating a lowering of the local symmetry~\cite{Lu2017}. This has been associated with changes in local charge distribution around Na ions which is surrounded by six OsO$_6$ octahedra and therefore directly reflects shifts of oxygen equilibrium positions. Since the time-scale of NMR experiments is much longer than $t_0$, in the dynamic JT scenario the local symmetry breaks only below $T_Q$, while in the static scenario the triplet line would have been observed at all temperatures, which is in clear contradiction with experimental results~\cite{Lu2017}. We note that the deformation due to the JT effect is significantly larger than what occurs below $T_Q$, therefore the splitting of the NMR line would have been clearly seen.

Finally, an argument against the random distribution of locally deformed octahedra comes from the consideration of the effect on local crystal environment. As revealed by high-resolution synchrotron experiments~\cite{Hirai2020}, below $T_Q$ the equilibrium of apical oxygen ions shifts by $\sim +0.36$\,\%, accompanied by a unit cell tetragonal elongation of $\sim +0.15$\,\%. Within the static JT scenario, the random orientation of local distortions would lead to a significant local strain on the surrounding cage of eight Ba ions (almost 1\,\% of local compression) in order to maintain the observed high-symmetry, cubic unit cell environment above $T_Q$. Since the quadrupolar moments are randomly frozen, and with such a high lattice strain, there is a clear lack of a feasible mechanism which would lead to the observed \textit{coherent} distortions below $T_Q$ along the $c$-axis, for both apical oxygen ions and the unit cell as a whole. Additionally, if the local distortions are static, the system could remove the strain by lowering the unit cell symmetry from cubic to tetragonal, which is the scenario observed in a series of JT-active $d^9$ double perovskites $A_2$Cu$B'$O$_6$ ($A =$ Ba, Sr; $B' =$ W, Te)~\cite{Iwanaga1999}.

The dynamic JT effect ensures the local and global cubic symmetry above $T_\mathrm{Q}$, in accordance with experimental observations in both \ce{Ba2MgReO6}~\cite{Hirai2020} and \ce{Ba2NaOsO6}~\cite{Lu2017}, as well as with \textit{ab initio} results reported for \ce{Ba2NaOsO6}~\cite{Iwahara2018}. Without the shared ligand between \ce{ReO6} octahedra, a feature characteristic for double perovskites, the dynamic distortions on one cluster are not correlated with its neighbours. Following the development of the charge quadrupolar order, where a \textit{coherent} charge redistribution occurs on Re ions, the PES becomes slightly renormalized which leads to a slight, but \textit{coherent} shift of the equilibrium position of oxygen ions, giving rise to the signal observed in REXS experiments~\cite{Hirai2020,Soh2023}. Crucially, the dynamic state of \ce{ReO6} octahedra continues to dominate the local Re environment even within the multi-polar ordered state. Even with a complete model of a spin-orbit-vibronic coupling the qualitative picture remains the same~\cite{Iwahara2023}.

The revelation that the ground-state doublet dominates the low temperature physics in this class of materials establishes a new paradigm from the thermodynamic point of view. With two phase transitions and two order parameters, it is puzzling how to reconcile $R \mathrm{ln}2$ across \textit{both} $T_\mathrm{M}$ and $T_\mathrm{Q}$. Concurrently, specific heat at $T_\mathrm{Q}$ does not exhibit a typical $\lambda$-type profile and, as we have shown, no lifting of degeneracy accompanies the quadrupolar order. Some of the entropy ($\sim 1/3R \mathrm{ln}2$) is released above $T_\mathrm{M}$, pointing to a reduction from a three-dimensional towards a two-dimensional phase space for fluctuating magnetic moments~\cite{Mansouri2021}. The plane of fluctuation below $T_\mathrm{Q}$ is then dictated by the orientation of quadrupolar moments $\mathcal{T}_x$ and $\mathcal{T}_z$, being perpendicular to the $\mathcal{T}_z$-induced elongation. At $T_\mathrm{M}$ the directional locking within that plane leads to magnetic dipole order and removes the remaining entropy. We infer that the magnetic and quadrupolar order parameters are part of a single, entangled object that goes through a two-step transition~\cite{Lovesey2021}.

Spin-orbit entanglement plays another crucial role that relates to the dynamical state of \ce{ReO6} octahedra. In the weak SOC limit, the available states forming a Mexican hat potential are composed of purely orbital contributions, each state allowing for two spin projections. For the strong SOC case involving 5$d$ orbitals, the entanglement entails a varying spin projection along the degeneracy line, resulting in a dynamically renormalized directional coupling between neighbouring \ce{ReO6} octahedra. Similar to Kitaev physics~\cite{Takagi2019}, and aided by the frustration on an FCC lattice, it could be envisaged that such a renormalization could lead to a spin-liquid state or other exotic quantum phases.

\textbf{Methods}

\textit{Sample preparation}

Single crystals were grown by the flux method following the protocol described in previous publications~\cite{Hirai2019,Hirai2020,Pasztorova2023}. The crystals exhibit octahedral morphology, forming flat triangles along [111] and equivalent directions. Specific heat measurements up to 14\,T were performed on as-grown crystals (up to 1\,mm in size), while $B = 35$\,T experiment was performed on a 0.3\,mm polished crystal. For the Re $L_3$ edge RIXS experiment the sample was glued on a copper sample holder with the [111] direction perpendicular to the surface. An in-situ cleaving has been utilized for the O $K$ edge RIXS experiment.

\textit{Resonant inelastic x-ray scattering}

The experiments at the Re $L_3$ edge were performed at BL11XU of SPring-8, Japan synchrotron facility. Incident x-rays were monochromatized by a Si(111) double-crystal monochromator and a two-bounce channel-cut monochromator Si(444), and $\pi$-polarized x-rays were irradiated on the samples. Horizontally scattered photons were energy-analyzed by a Ge(733) analyzer. The total energy resolution was 140\,meV (full width at half maximum). The experiments at the O $K$ edge were performed at ADRESS beamline of the Swiss light source (SLS) at the Paul Scherrer Institute (PSI), Switzerland in both the normal and grazing configuration with 50$^\circ$ between incoming and outgoing beams~\cite{Strocov2010}.

\textit{Embedding quantum chemistry calculations}

Many-body wavefunction calculations on electrostatically embedded finite-size clusters were performed using the Molpro package~\cite{Molpro}. The model involved a 21-atom cluster consisting of the \ce{ReO6} octahedron together with the nearest 6 Mg and 8 Ba atoms. Re atoms were represented with core potentials and triple-zeta plus two polarization $f$-functions basis functions~\cite{ReBasis}. All-electron triple-zeta basis functions were used for O atoms~\cite{OBasis}. Both Mg and Ba atoms in the cluster were described by effective core potential and supplemented with a single $s$ basis function ~\cite{MgBasis1,MgBasis2}. The effect of the crystal lattice surrounding the cluster was treated at the level of Madelung ionic potential, as described in Ref.~\cite{Ewald}. For the complete active space self-consistent field (CASSCF) calculations, an active space of $3$~$t_{2g} + 2$~$e_g$ orbitals was used. The optimisation was carried out for an average of 5 states, each of unit weight, of the scalar relativistic Hamiltonian. Multireference configuration interaction (MRCI) treatment was performed with single and double substitutions with respect to the CASSCF reference, as described in Refs.~\cite{MRCI1,MRCI2}. The spin–orbit coupling correction was added as described in Ref.~\cite{SOC}. The PES was mapped out with respect to isotropic mode $\boldsymbol{Q}_{iso} = (x_1 - x_2 + y_3 - y_4 + z_5 - z_6)$, and $e_g$-type distortions $\boldsymbol{Q_2} = (-x_1 + x_2 + y_3 - y_4)$ and $\boldsymbol{Q_3} = (-x_1 +x_2 -y_3 + y_4+ 2z_5- 2z_6)$, with superscripts 1,2 referring to the $x$-axis O atoms, 3,4 to the $y$-axis O atoms and 5,6 to the $z$-axis O atoms. 

In order to asses the PES, calculations were performed on a grid that varies isotropic distortions were varied from 0.5\% to $-$1.5\% with 0.5\% steps, with amplitudes referring to non-normalised $\boldsymbol{Q}_{iso}=[1,1,1]$ and the percentage being relative to the undistorted \SI{1.926}{\angstrom} Re$-$O interatomic distance. A \ang{60} sector of the $Q_2$$-$$Q_3$ plane was simulated, with angle being varied with steps of \ang{15} and radius being varied from 0 to 3.5\%, with step of 0.25\%, with amplitudes referring to normalized $\boldsymbol{Q_2}=\frac{1}{\sqrt{2}}[-1,1,0]$ and $\boldsymbol{Q_3}=\frac{1}{\sqrt{6}}[1,1,-2]$. The stability of the minimum with respect to the $t_{2g}$-type distortions $\boldsymbol{Q_4} = (y_1 - y_2 + x_3 - x_4)$ was confirmed. In addition, an attempt was made to find a minimum with a non-zero amplitude of the $t_{2g}$-type distortions by alternating ${Q}_{iso}$, ${Q}_{2}$, ${Q}_{3}$ and ${Q}_{4}$ as energy optimization coordinates, starting at the cubic configuration. No such minimum was found, with all paths considered converging back to the minimum described in the article.


\textit{Specific heat}

For magnetic fields up to 14\,T we used a PPMS system (Quantum Design) that utilizes a relaxation method. The measurement at 35\,T was performed at LNCMI, Grenoble, France using a Cernox thermometer as a platform and a semi-adiabatic method. In both cases the addenda has been measured separately and subtracted to obtain specific heat of the sample. Zero-field measurements have been used to normalize the results from two setups.

\textbf{Acknowledgments}.

H.M.R. acknowledges the funding by the Swiss National Science Foundation (SNF) Quantum Magnetism grant (No. 200020-188648).  Re $L_3$ edge RIXS experiments were supported by ``Advanced Research Infrastructure for Materials and Nanotechnology in Japan (ARIM)" of the Ministry of Education, Culture, Sports, Science and Technology (MEXT), Japan (Proposal No. JPMXP1222QS0107) and were performed at the QST experimental station at BL11XU of SPring-8, with the approval of the Japan Synchrotron Radiation Research Institute (JASRI) (SPring-8 Proposal No. 2022B3596). The O $K$-edge XAS and RIXS experiments were performed at the ADRESS beamline of the Swiss Light Source at the Paul Scherrer Institut (PSI). H.M.R. and J.R.S acknowledges the funding from the European Research Council (ERC) under the European Union’s Horizon 2020 research and innovation program projects HERO (Grant No. 810451).  The experimental work at PSI is supported by the Swiss National Science Foundation through project nos. 178867 and 207904. T.Y. is funded by a CROSS project of PSI. Y.W. and G.C.W. acknowledge funding from the European Union's Horizon 2020 research and innovation programme under the Marie Sklodowska-Curie grant agreement No. 884104 (PSI-FELLOW-II-3i program). We acknowledge the support from LNCMI-CNRS, a member of the European Magnetic Field Laboratory (EMFL).

\textbf{Author contributions}.

Single crystals were grown by D.H., D.T. and J.P. RIXS data was measured and analyzed by I.\v{Z}., J-R.S., F.P., K.I., Y.W., W.Z., C.G., T.Y., T.S. and H.M.R. Embedding quantum chemistry calculations were performed by O.M. and R.Y. under guidance of O.V.Y. and H.M.R. Specific heat was measured and analyzed by I.\v{Z}. and A.D. The phonon contribution to specific heat was calculated by A.M.T. The manuscript was written by I.\v{Z}. with contributions from all co-authors.

\bibliography{referencesBMRO.bib}

\begin{thebibliography}{62}%
\makeatletter
\providecommand \@ifxundefined [1]{%
 \@ifx{#1\undefined}
}%
\providecommand \@ifnum [1]{%
 \ifnum #1\expandafter \@firstoftwo
 \else \expandafter \@secondoftwo
 \fi
}%
\providecommand \@ifx [1]{%
 \ifx #1\expandafter \@firstoftwo
 \else \expandafter \@secondoftwo
 \fi
}%
\providecommand \natexlab [1]{#1}%
\providecommand \enquote  [1]{``#1''}%
\providecommand \bibnamefont  [1]{#1}%
\providecommand \bibfnamefont [1]{#1}%
\providecommand \citenamefont [1]{#1}%
\providecommand \href@noop [0]{\@secondoftwo}%
\providecommand \href [0]{\begingroup \@sanitize@url \@href}%
\providecommand \@href[1]{\@@startlink{#1}\@@href}%
\providecommand \@@href[1]{\endgroup#1\@@endlink}%
\providecommand \@sanitize@url [0]{\catcode `\\12\catcode `\$12\catcode `\&12\catcode `\#12\catcode `\^12\catcode `\_12\catcode `\%12\relax}%
\providecommand \@@startlink[1]{}%
\providecommand \@@endlink[0]{}%
\providecommand \url  [0]{\begingroup\@sanitize@url \@url }%
\providecommand \@url [1]{\endgroup\@href {#1}{\urlprefix }}%
\providecommand \urlprefix  [0]{URL }%
\providecommand \Eprint [0]{\href }%
\providecommand \doibase [0]{https://doi.org/}%
\providecommand \selectlanguage [0]{\@gobble}%
\providecommand \bibinfo  [0]{\@secondoftwo}%
\providecommand \bibfield  [0]{\@secondoftwo}%
\providecommand \translation [1]{[#1]}%
\providecommand \BibitemOpen [0]{}%
\providecommand \bibitemStop [0]{}%
\providecommand \bibitemNoStop [0]{.\EOS\space}%
\providecommand \EOS [0]{\spacefactor3000\relax}%
\providecommand \BibitemShut  [1]{\csname bibitem#1\endcsname}%
\let\auto@bib@innerbib\@empty
\bibitem [{\citenamefont {Bersuker}(2006)}]{Bersuker2006}%
  \BibitemOpen
  \bibfield  {author} {\bibinfo {author} {\bibfnamefont {I.}~\bibnamefont {Bersuker}},\ }\href {https://doi.org/10.1017/CBO9780511524769} {\emph {\bibinfo {title} {{The Jahn-Teller Effect}}}}\ (\bibinfo  {publisher} {Cambridge University Press},\ \bibinfo {year} {2006})\BibitemShut {NoStop}%
\bibitem [{\citenamefont {Li}\ \emph {et~al.}(2021)\citenamefont {Li}, \citenamefont {Zhang}, \citenamefont {Vendrell}, \citenamefont {Guo}, \citenamefont {Zhu}, \citenamefont {Gao}, \citenamefont {Cao}, \citenamefont {Guo}, \citenamefont {Su}, \citenamefont {Cao}, \citenamefont {Luo}, \citenamefont {Yan}, \citenamefont {Zhou}, \citenamefont {Liu}, \citenamefont {Li},\ and\ \citenamefont {Lu}}]{Li2021}%
  \BibitemOpen
  \bibfield  {author} {\bibinfo {author} {\bibfnamefont {M.}~\bibnamefont {Li}}, \bibinfo {author} {\bibfnamefont {M.}~\bibnamefont {Zhang}}, \bibinfo {author} {\bibfnamefont {O.}~\bibnamefont {Vendrell}}, \bibinfo {author} {\bibfnamefont {Z.}~\bibnamefont {Guo}}, \bibinfo {author} {\bibfnamefont {Q.}~\bibnamefont {Zhu}}, \bibinfo {author} {\bibfnamefont {X.}~\bibnamefont {Gao}}, \bibinfo {author} {\bibfnamefont {L.}~\bibnamefont {Cao}}, \bibinfo {author} {\bibfnamefont {K.}~\bibnamefont {Guo}}, \bibinfo {author} {\bibfnamefont {Q.-Q.}\ \bibnamefont {Su}}, \bibinfo {author} {\bibfnamefont {W.}~\bibnamefont {Cao}}, \bibinfo {author} {\bibfnamefont {S.}~\bibnamefont {Luo}}, \bibinfo {author} {\bibfnamefont {J.}~\bibnamefont {Yan}}, \bibinfo {author} {\bibfnamefont {Y.}~\bibnamefont {Zhou}}, \bibinfo {author} {\bibfnamefont {Y.}~\bibnamefont {Liu}}, \bibinfo {author} {\bibfnamefont {Z.}~\bibnamefont {Li}},\ and\ \bibinfo {author} {\bibfnamefont {P.}~\bibnamefont {Lu}},\ }\bibfield  {title} {\bibinfo {title}
  {{Ultrafast imaging of spontaneous symmetry breaking in a photoionized molecular system}},\ }\href {https://doi.org/10.1038/s41467-021-24309-z} {\bibfield  {journal} {\bibinfo  {journal} {Nature Communications}\ }\textbf {\bibinfo {volume} {12}},\ \bibinfo {pages} {4233} (\bibinfo {year} {2021})}\BibitemShut {NoStop}%
\bibitem [{\citenamefont {Ridente}\ \emph {et~al.}(2023)\citenamefont {Ridente}, \citenamefont {Hait}, \citenamefont {Haugen}, \citenamefont {Ross}, \citenamefont {Neumark}, \citenamefont {Head-Gordon},\ and\ \citenamefont {Leone}}]{Ridente2023}%
  \BibitemOpen
  \bibfield  {author} {\bibinfo {author} {\bibfnamefont {E.}~\bibnamefont {Ridente}}, \bibinfo {author} {\bibfnamefont {D.}~\bibnamefont {Hait}}, \bibinfo {author} {\bibfnamefont {E.~A.}\ \bibnamefont {Haugen}}, \bibinfo {author} {\bibfnamefont {A.~D.}\ \bibnamefont {Ross}}, \bibinfo {author} {\bibfnamefont {D.~M.}\ \bibnamefont {Neumark}}, \bibinfo {author} {\bibfnamefont {M.}~\bibnamefont {Head-Gordon}},\ and\ \bibinfo {author} {\bibfnamefont {S.~R.}\ \bibnamefont {Leone}},\ }\bibfield  {title} {\bibinfo {title} {{Femtosecond symmetry breaking and coherent relaxation of methane cations via x-ray spectroscopy}},\ }\href {https://doi.org/10.1126/science.adg4421} {\bibfield  {journal} {\bibinfo  {journal} {Science}\ }\textbf {\bibinfo {volume} {380}},\ \bibinfo {pages} {713} (\bibinfo {year} {2023})},\ \Eprint {https://arxiv.org/abs/https://www.science.org/doi/pdf/10.1126/science.adg4421} {https://www.science.org/doi/pdf/10.1126/science.adg4421} \BibitemShut {NoStop}%
\bibitem [{\citenamefont {Barlow}\ \emph {et~al.}(2022)\citenamefont {Barlow}, \citenamefont {Eng}, \citenamefont {Ivalo}, \citenamefont {Coletta}, \citenamefont {Brechin}, \citenamefont {Penfold},\ and\ \citenamefont {Johansson}}]{Barlow2022}%
  \BibitemOpen
  \bibfield  {author} {\bibinfo {author} {\bibfnamefont {K.}~\bibnamefont {Barlow}}, \bibinfo {author} {\bibfnamefont {J.}~\bibnamefont {Eng}}, \bibinfo {author} {\bibfnamefont {I.}~\bibnamefont {Ivalo}}, \bibinfo {author} {\bibfnamefont {M.}~\bibnamefont {Coletta}}, \bibinfo {author} {\bibfnamefont {E.~K.}\ \bibnamefont {Brechin}}, \bibinfo {author} {\bibfnamefont {T.~J.}\ \bibnamefont {Penfold}},\ and\ \bibinfo {author} {\bibfnamefont {J.~O.}\ \bibnamefont {Johansson}},\ }\bibfield  {title} {\bibinfo {title} {{Photoinduced Jahn–Teller switch in Mn(iii) terpyridine complexes}},\ }\href {https://doi.org/10.1039/D2DT00889K} {\bibfield  {journal} {\bibinfo  {journal} {Dalton Trans.}\ }\textbf {\bibinfo {volume} {51}},\ \bibinfo {pages} {10751} (\bibinfo {year} {2022})}\BibitemShut {NoStop}%
\bibitem [{\citenamefont {Babar}\ and\ \citenamefont {Kabir}(2018)}]{Babar2018}%
  \BibitemOpen
  \bibfield  {author} {\bibinfo {author} {\bibfnamefont {R.}~\bibnamefont {Babar}}\ and\ \bibinfo {author} {\bibfnamefont {M.}~\bibnamefont {Kabir}},\ }\bibfield  {title} {\bibinfo {title} {{Gate-dependent vacancy diffusion in graphene}},\ }\href {https://doi.org/10.1103/PhysRevB.98.075439} {\bibfield  {journal} {\bibinfo  {journal} {Phys. Rev. B}\ }\textbf {\bibinfo {volume} {98}},\ \bibinfo {pages} {075439} (\bibinfo {year} {2018})}\BibitemShut {NoStop}%
\bibitem [{\citenamefont {Keller}\ \emph {et~al.}(2008)\citenamefont {Keller}, \citenamefont {Bussmann-Holder},\ and\ \citenamefont {Müller}}]{Keller2008}%
  \BibitemOpen
  \bibfield  {author} {\bibinfo {author} {\bibfnamefont {H.}~\bibnamefont {Keller}}, \bibinfo {author} {\bibfnamefont {A.}~\bibnamefont {Bussmann-Holder}},\ and\ \bibinfo {author} {\bibfnamefont {K.~A.}\ \bibnamefont {Müller}},\ }\bibfield  {title} {\bibinfo {title} {{Jahn–Teller physics and high-Tc superconductivity}},\ }\href {https://doi.org/https://doi.org/10.1016/S1369-7021(08)70178-0} {\bibfield  {journal} {\bibinfo  {journal} {Materials Today}\ }\textbf {\bibinfo {volume} {11}},\ \bibinfo {pages} {38} (\bibinfo {year} {2008})}\BibitemShut {NoStop}%
\bibitem [{\citenamefont {Zheng}\ \emph {et~al.}(2003)\citenamefont {Zheng}, \citenamefont {Li}, \citenamefont {Tang}, \citenamefont {Yang}, \citenamefont {Wang}, \citenamefont {Li}, \citenamefont {Wang},\ and\ \citenamefont {Ku}}]{Zheng2003}%
  \BibitemOpen
  \bibfield  {author} {\bibinfo {author} {\bibfnamefont {R.~K.}\ \bibnamefont {Zheng}}, \bibinfo {author} {\bibfnamefont {G.}~\bibnamefont {Li}}, \bibinfo {author} {\bibfnamefont {A.~N.}\ \bibnamefont {Tang}}, \bibinfo {author} {\bibfnamefont {Y.}~\bibnamefont {Yang}}, \bibinfo {author} {\bibfnamefont {W.}~\bibnamefont {Wang}}, \bibinfo {author} {\bibfnamefont {X.~G.}\ \bibnamefont {Li}}, \bibinfo {author} {\bibfnamefont {Z.~D.}\ \bibnamefont {Wang}},\ and\ \bibinfo {author} {\bibfnamefont {H.~C.}\ \bibnamefont {Ku}},\ }\bibfield  {title} {\bibinfo {title} {{The role of the cooperative Jahn–Teller effect in the charge-ordered ${\mathrm{La}}_{1-x}{\mathrm{Ca}}_{x}{\mathrm{Mn}}{\mathrm{O}}_3 (0.5 x 0.87)$ manganites}},\ }\href {https://doi.org/10.1063/1.1635662} {\bibfield  {journal} {\bibinfo  {journal} {Applied Physics Letters}\ }\textbf {\bibinfo {volume} {83}},\ \bibinfo {pages} {5250} (\bibinfo {year} {2003})},\ \Eprint
  {https://arxiv.org/abs/https://pubs.aip.org/aip/apl/article-pdf/83/25/5250/8783891/5250\_1\_online.pdf} {https://pubs.aip.org/aip/apl/article-pdf/83/25/5250/8783891/5250\_1\_online.pdf} \BibitemShut {NoStop}%
\bibitem [{\citenamefont {Vitalone}\ \emph {et~al.}(2022)\citenamefont {Vitalone}, \citenamefont {Sternbach}, \citenamefont {Foutty}, \citenamefont {McLeod}, \citenamefont {Sow}, \citenamefont {Golez}, \citenamefont {Nakamura}, \citenamefont {Maeno}, \citenamefont {Pasupathy}, \citenamefont {Georges}, \citenamefont {Millis},\ and\ \citenamefont {Basov}}]{Vitalone2022}%
  \BibitemOpen
  \bibfield  {author} {\bibinfo {author} {\bibfnamefont {R.~A.}\ \bibnamefont {Vitalone}}, \bibinfo {author} {\bibfnamefont {A.~J.}\ \bibnamefont {Sternbach}}, \bibinfo {author} {\bibfnamefont {B.~A.}\ \bibnamefont {Foutty}}, \bibinfo {author} {\bibfnamefont {A.~S.}\ \bibnamefont {McLeod}}, \bibinfo {author} {\bibfnamefont {C.}~\bibnamefont {Sow}}, \bibinfo {author} {\bibfnamefont {D.}~\bibnamefont {Golez}}, \bibinfo {author} {\bibfnamefont {F.}~\bibnamefont {Nakamura}}, \bibinfo {author} {\bibfnamefont {Y.}~\bibnamefont {Maeno}}, \bibinfo {author} {\bibfnamefont {A.~N.}\ \bibnamefont {Pasupathy}}, \bibinfo {author} {\bibfnamefont {A.}~\bibnamefont {Georges}}, \bibinfo {author} {\bibfnamefont {A.~J.}\ \bibnamefont {Millis}},\ and\ \bibinfo {author} {\bibfnamefont {D.~N.}\ \bibnamefont {Basov}},\ }\bibfield  {title} {\bibinfo {title} {{Nanoscale Femtosecond Dynamics of Mott Insulator (Ca$_{0.99}$Sr$_{0.01}$)$_2$RuO$_4$}},\ }\href {https://doi.org/10.1021/acs.nanolett.2c00581} {\bibfield  {journal} {\bibinfo
  {journal} {Nano Letters}\ }\textbf {\bibinfo {volume} {22}},\ \bibinfo {pages} {5689} (\bibinfo {year} {2022})}\BibitemShut {NoStop}%
\bibitem [{\citenamefont {Geirhos}\ \emph {et~al.}(2021)\citenamefont {Geirhos}, \citenamefont {Langmann}, \citenamefont {Prodan}, \citenamefont {Tsirlin}, \citenamefont {Missiul}, \citenamefont {Eickerling}, \citenamefont {Jesche}, \citenamefont {Tsurkan}, \citenamefont {Lunkenheimer}, \citenamefont {Scherer},\ and\ \citenamefont {K\'ezsm\'arki}}]{Geirhos2021}%
  \BibitemOpen
  \bibfield  {author} {\bibinfo {author} {\bibfnamefont {K.}~\bibnamefont {Geirhos}}, \bibinfo {author} {\bibfnamefont {J.}~\bibnamefont {Langmann}}, \bibinfo {author} {\bibfnamefont {L.}~\bibnamefont {Prodan}}, \bibinfo {author} {\bibfnamefont {A.~A.}\ \bibnamefont {Tsirlin}}, \bibinfo {author} {\bibfnamefont {A.}~\bibnamefont {Missiul}}, \bibinfo {author} {\bibfnamefont {G.}~\bibnamefont {Eickerling}}, \bibinfo {author} {\bibfnamefont {A.}~\bibnamefont {Jesche}}, \bibinfo {author} {\bibfnamefont {V.}~\bibnamefont {Tsurkan}}, \bibinfo {author} {\bibfnamefont {P.}~\bibnamefont {Lunkenheimer}}, \bibinfo {author} {\bibfnamefont {W.}~\bibnamefont {Scherer}},\ and\ \bibinfo {author} {\bibfnamefont {I.}~\bibnamefont {K\'ezsm\'arki}},\ }\bibfield  {title} {\bibinfo {title} {{Cooperative Cluster Jahn-Teller Effect as a Possible Route to Antiferroelectricity}},\ }\href {https://doi.org/10.1103/PhysRevLett.126.187601} {\bibfield  {journal} {\bibinfo  {journal} {Phys. Rev. Lett.}\ }\textbf {\bibinfo {volume} {126}},\
  \bibinfo {pages} {187601} (\bibinfo {year} {2021})}\BibitemShut {NoStop}%
\bibitem [{\citenamefont {Khomskii}\ and\ \citenamefont {Streltsov}(2021)}]{Khomskii2021}%
  \BibitemOpen
  \bibfield  {author} {\bibinfo {author} {\bibfnamefont {D.~I.}\ \bibnamefont {Khomskii}}\ and\ \bibinfo {author} {\bibfnamefont {S.~V.}\ \bibnamefont {Streltsov}},\ }\bibfield  {title} {\bibinfo {title} {{Orbital Effects in Solids: Basics, Recent Progress, and Opportunities}},\ }\href {https://doi.org/10.1021/acs.chemrev.0c00579} {\bibfield  {journal} {\bibinfo  {journal} {Chemical Reviews}\ }\textbf {\bibinfo {volume} {121}},\ \bibinfo {pages} {2992} (\bibinfo {year} {2021})}\BibitemShut {NoStop}%
\bibitem [{\citenamefont {Millis}\ \emph {et~al.}(1996)\citenamefont {Millis}, \citenamefont {Shraiman},\ and\ \citenamefont {Mueller}}]{Millis1996}%
  \BibitemOpen
  \bibfield  {author} {\bibinfo {author} {\bibfnamefont {A.~J.}\ \bibnamefont {Millis}}, \bibinfo {author} {\bibfnamefont {B.~I.}\ \bibnamefont {Shraiman}},\ and\ \bibinfo {author} {\bibfnamefont {R.}~\bibnamefont {Mueller}},\ }\bibfield  {title} {\bibinfo {title} {{Dynamic Jahn-Teller Effect and Colossal Magnetoresistance in ${\mathrm{La}}_{1\ensuremath{-}\mathit{x}}{\mathrm{Sr}}_{\mathit{x}}{\mathrm{MnO}}_{3}$}},\ }\href {https://doi.org/10.1103/PhysRevLett.77.175} {\bibfield  {journal} {\bibinfo  {journal} {Phys. Rev. Lett.}\ }\textbf {\bibinfo {volume} {77}},\ \bibinfo {pages} {175} (\bibinfo {year} {1996})}\BibitemShut {NoStop}%
\bibitem [{\citenamefont {Kayanuma}\ and\ \citenamefont {Nakamura}(2017)}]{Kayanuma2017}%
  \BibitemOpen
  \bibfield  {author} {\bibinfo {author} {\bibfnamefont {Y.}~\bibnamefont {Kayanuma}}\ and\ \bibinfo {author} {\bibfnamefont {K.~G.}\ \bibnamefont {Nakamura}},\ }\bibfield  {title} {\bibinfo {title} {{Dynamic Jahn-Teller viewpoint for generation mechanism of asymmetric modes of coherent phonons}},\ }\href {https://doi.org/10.1103/PhysRevB.95.104302} {\bibfield  {journal} {\bibinfo  {journal} {Phys. Rev. B}\ }\textbf {\bibinfo {volume} {95}},\ \bibinfo {pages} {104302} (\bibinfo {year} {2017})}\BibitemShut {NoStop}%
\bibitem [{\citenamefont {Ribeiro}\ and\ \citenamefont {Yuen-Zhou}(2018)}]{Ribeiro2018}%
  \BibitemOpen
  \bibfield  {author} {\bibinfo {author} {\bibfnamefont {R.~F.}\ \bibnamefont {Ribeiro}}\ and\ \bibinfo {author} {\bibfnamefont {J.}~\bibnamefont {Yuen-Zhou}},\ }\bibfield  {title} {\bibinfo {title} {{Continuous vibronic symmetries in Jahn–Teller models}},\ }\href {https://doi.org/10.1088/1361-648X/aac89e} {\bibfield  {journal} {\bibinfo  {journal} {Journal of Physics: Condensed Matter}\ }\textbf {\bibinfo {volume} {30}},\ \bibinfo {pages} {333001} (\bibinfo {year} {2018})}\BibitemShut {NoStop}%
\bibitem [{\citenamefont {Klupp}\ \emph {et~al.}(2012)\citenamefont {Klupp}, \citenamefont {Matus}, \citenamefont {Kamar{\'a}s}, \citenamefont {Ganin}, \citenamefont {McLennan}, \citenamefont {Rosseinsky}, \citenamefont {Takabayashi}, \citenamefont {McDonald},\ and\ \citenamefont {Prassides}}]{Klupp2012}%
  \BibitemOpen
  \bibfield  {author} {\bibinfo {author} {\bibfnamefont {G.}~\bibnamefont {Klupp}}, \bibinfo {author} {\bibfnamefont {P.}~\bibnamefont {Matus}}, \bibinfo {author} {\bibfnamefont {K.}~\bibnamefont {Kamar{\'a}s}}, \bibinfo {author} {\bibfnamefont {A.~Y.}\ \bibnamefont {Ganin}}, \bibinfo {author} {\bibfnamefont {A.}~\bibnamefont {McLennan}}, \bibinfo {author} {\bibfnamefont {M.~J.}\ \bibnamefont {Rosseinsky}}, \bibinfo {author} {\bibfnamefont {Y.}~\bibnamefont {Takabayashi}}, \bibinfo {author} {\bibfnamefont {M.~T.}\ \bibnamefont {McDonald}},\ and\ \bibinfo {author} {\bibfnamefont {K.}~\bibnamefont {Prassides}},\ }\bibfield  {title} {\bibinfo {title} {{Dynamic Jahn--Teller effect in the parent insulating state of the molecular superconductor Cs$_3$C$_{60}$}},\ }\href {https://doi.org/10.1038/ncomms1910} {\bibfield  {journal} {\bibinfo  {journal} {Nature Communications}\ }\textbf {\bibinfo {volume} {3}},\ \bibinfo {pages} {912} (\bibinfo {year} {2012})}\BibitemShut {NoStop}%
\bibitem [{\citenamefont {Iwahara}\ and\ \citenamefont {Chibotaru}(2013)}]{Iwahara2013}%
  \BibitemOpen
  \bibfield  {author} {\bibinfo {author} {\bibfnamefont {N.}~\bibnamefont {Iwahara}}\ and\ \bibinfo {author} {\bibfnamefont {L.~F.}\ \bibnamefont {Chibotaru}},\ }\bibfield  {title} {\bibinfo {title} {{Dynamical Jahn-Teller Effect and Antiferromagnetism in ${\mathrm{Cs}}_{3}{\mathrm{C}}_{60}$}},\ }\href {https://doi.org/10.1103/PhysRevLett.111.056401} {\bibfield  {journal} {\bibinfo  {journal} {Phys. Rev. Lett.}\ }\textbf {\bibinfo {volume} {111}},\ \bibinfo {pages} {056401} (\bibinfo {year} {2013})}\BibitemShut {NoStop}%
\bibitem [{\citenamefont {Zadik}\ \emph {et~al.}(2015)\citenamefont {Zadik}, \citenamefont {Takabayashi}, \citenamefont {Klupp}, \citenamefont {Colman}, \citenamefont {Ganin}, \citenamefont {Potočnik}, \citenamefont {Jeglič}, \citenamefont {Arčon}, \citenamefont {Matus}, \citenamefont {Kamarás}, \citenamefont {Kasahara}, \citenamefont {Iwasa}, \citenamefont {Fitch}, \citenamefont {Ohishi}, \citenamefont {Garbarino}, \citenamefont {Kato}, \citenamefont {Rosseinsky},\ and\ \citenamefont {Prassides}}]{Zadik2015}%
  \BibitemOpen
  \bibfield  {author} {\bibinfo {author} {\bibfnamefont {R.~H.}\ \bibnamefont {Zadik}}, \bibinfo {author} {\bibfnamefont {Y.}~\bibnamefont {Takabayashi}}, \bibinfo {author} {\bibfnamefont {G.}~\bibnamefont {Klupp}}, \bibinfo {author} {\bibfnamefont {R.~H.}\ \bibnamefont {Colman}}, \bibinfo {author} {\bibfnamefont {A.~Y.}\ \bibnamefont {Ganin}}, \bibinfo {author} {\bibfnamefont {A.}~\bibnamefont {Potočnik}}, \bibinfo {author} {\bibfnamefont {P.}~\bibnamefont {Jeglič}}, \bibinfo {author} {\bibfnamefont {D.}~\bibnamefont {Arčon}}, \bibinfo {author} {\bibfnamefont {P.}~\bibnamefont {Matus}}, \bibinfo {author} {\bibfnamefont {K.}~\bibnamefont {Kamarás}}, \bibinfo {author} {\bibfnamefont {Y.}~\bibnamefont {Kasahara}}, \bibinfo {author} {\bibfnamefont {Y.}~\bibnamefont {Iwasa}}, \bibinfo {author} {\bibfnamefont {A.~N.}\ \bibnamefont {Fitch}}, \bibinfo {author} {\bibfnamefont {Y.}~\bibnamefont {Ohishi}}, \bibinfo {author} {\bibfnamefont {G.}~\bibnamefont {Garbarino}}, \bibinfo {author} {\bibfnamefont
  {K.}~\bibnamefont {Kato}}, \bibinfo {author} {\bibfnamefont {M.~J.}\ \bibnamefont {Rosseinsky}},\ and\ \bibinfo {author} {\bibfnamefont {K.}~\bibnamefont {Prassides}},\ }\bibfield  {title} {\bibinfo {title} {{Optimized unconventional superconductivity in a molecular Jahn-Teller metal}},\ }\href {https://doi.org/10.1126/sciadv.1500059} {\bibfield  {journal} {\bibinfo  {journal} {Science Advances}\ }\textbf {\bibinfo {volume} {1}},\ \bibinfo {pages} {e1500059} (\bibinfo {year} {2015})}\BibitemShut {NoStop}%
\bibitem [{\citenamefont {Wieczorek}\ \emph {et~al.}(2006)\citenamefont {Wieczorek}, \citenamefont {Ziebiniska}, \citenamefont {Ujma}, \citenamefont {Szot}, \citenamefont {Gorny}, \citenamefont {Franke}, \citenamefont {Koperski}, \citenamefont {Soszynski},\ and\ \citenamefont {Roleder}}]{Wieczorek2006}%
  \BibitemOpen
  \bibfield  {author} {\bibinfo {author} {\bibfnamefont {K.}~\bibnamefont {Wieczorek}}, \bibinfo {author} {\bibfnamefont {A.}~\bibnamefont {Ziebiniska}}, \bibinfo {author} {\bibfnamefont {Z.}~\bibnamefont {Ujma}}, \bibinfo {author} {\bibfnamefont {K.}~\bibnamefont {Szot}}, \bibinfo {author} {\bibfnamefont {M.}~\bibnamefont {Gorny}}, \bibinfo {author} {\bibfnamefont {I.}~\bibnamefont {Franke}}, \bibinfo {author} {\bibfnamefont {J.}~\bibnamefont {Koperski}}, \bibinfo {author} {\bibfnamefont {A.}~\bibnamefont {Soszynski}},\ and\ \bibinfo {author} {\bibfnamefont {K.}~\bibnamefont {Roleder}},\ }\bibfield  {title} {\bibinfo {title} {{Electrostrictive and Piezoelectric Effect in BaTiO3 and PbZrO3 }},\ }\href {https://doi.org/10.1080/00150190600695743} {\bibfield  {journal} {\bibinfo  {journal} {Ferroelectrics}\ }\textbf {\bibinfo {volume} {336}},\ \bibinfo {pages} {61} (\bibinfo {year} {2006})}\BibitemShut {NoStop}%
\bibitem [{\citenamefont {Bersuker}(2015)}]{Bersuker2015}%
  \BibitemOpen
  \bibfield  {author} {\bibinfo {author} {\bibfnamefont {I.~B.}\ \bibnamefont {Bersuker}},\ }\bibfield  {title} {\bibinfo {title} {{Giant permittivity and electrostriction induced by dynamic Jahn-Teller and pseudo Jahn-Teller effects}},\ }\href {https://doi.org/10.1063/1.4936190} {\bibfield  {journal} {\bibinfo  {journal} {Applied Physics Letters}\ }\textbf {\bibinfo {volume} {107}},\ \bibinfo {pages} {202904} (\bibinfo {year} {2015})}\BibitemShut {NoStop}%
\bibitem [{\citenamefont {Hasan}\ and\ \citenamefont {Kane}(2010)}]{Hasan2010}%
  \BibitemOpen
  \bibfield  {author} {\bibinfo {author} {\bibfnamefont {M.~Z.}\ \bibnamefont {Hasan}}\ and\ \bibinfo {author} {\bibfnamefont {C.~L.}\ \bibnamefont {Kane}},\ }\bibfield  {title} {\bibinfo {title} {{Colloquium: Topological insulators}},\ }\href {https://doi.org/10.1103/RevModPhys.82.3045} {\bibfield  {journal} {\bibinfo  {journal} {Rev. Mod. Phys.}\ }\textbf {\bibinfo {volume} {82}},\ \bibinfo {pages} {3045} (\bibinfo {year} {2010})}\BibitemShut {NoStop}%
\bibitem [{\citenamefont {Armitage}\ \emph {et~al.}(2018)\citenamefont {Armitage}, \citenamefont {Mele},\ and\ \citenamefont {Vishwanath}}]{Armitage2018}%
  \BibitemOpen
  \bibfield  {author} {\bibinfo {author} {\bibfnamefont {N.~P.}\ \bibnamefont {Armitage}}, \bibinfo {author} {\bibfnamefont {E.~J.}\ \bibnamefont {Mele}},\ and\ \bibinfo {author} {\bibfnamefont {A.}~\bibnamefont {Vishwanath}},\ }\bibfield  {title} {\bibinfo {title} {{Weyl and Dirac semimetals in three-dimensional solids}},\ }\href {https://doi.org/10.1103/RevModPhys.90.015001} {\bibfield  {journal} {\bibinfo  {journal} {Rev. Mod. Phys.}\ }\textbf {\bibinfo {volume} {90}},\ \bibinfo {pages} {015001} (\bibinfo {year} {2018})}\BibitemShut {NoStop}%
\bibitem [{\citenamefont {Takagi}\ \emph {et~al.}(2019)\citenamefont {Takagi}, \citenamefont {Takayama}, \citenamefont {Jackeli}, \citenamefont {Khaliullin},\ and\ \citenamefont {Nagler}}]{Takagi2019}%
  \BibitemOpen
  \bibfield  {author} {\bibinfo {author} {\bibfnamefont {H.}~\bibnamefont {Takagi}}, \bibinfo {author} {\bibfnamefont {T.}~\bibnamefont {Takayama}}, \bibinfo {author} {\bibfnamefont {G.}~\bibnamefont {Jackeli}}, \bibinfo {author} {\bibfnamefont {G.}~\bibnamefont {Khaliullin}},\ and\ \bibinfo {author} {\bibfnamefont {S.~E.}\ \bibnamefont {Nagler}},\ }\bibfield  {title} {\bibinfo {title} {{Concept and realization of Kitaev quantum spin liquids}},\ }\href {https://doi.org/10.1038/s42254-019-0038-2} {\bibfield  {journal} {\bibinfo  {journal} {Nature Reviews Physics}\ }\textbf {\bibinfo {volume} {1}},\ \bibinfo {pages} {264} (\bibinfo {year} {2019})}\BibitemShut {NoStop}%
\bibitem [{\citenamefont {Plotnikova}\ \emph {et~al.}(2016)\citenamefont {Plotnikova}, \citenamefont {Daghofer}, \citenamefont {van~den Brink},\ and\ \citenamefont {Wohlfeld}}]{Plotnikova2016}%
  \BibitemOpen
  \bibfield  {author} {\bibinfo {author} {\bibfnamefont {E.~M.}\ \bibnamefont {Plotnikova}}, \bibinfo {author} {\bibfnamefont {M.}~\bibnamefont {Daghofer}}, \bibinfo {author} {\bibfnamefont {J.}~\bibnamefont {van~den Brink}},\ and\ \bibinfo {author} {\bibfnamefont {K.}~\bibnamefont {Wohlfeld}},\ }\bibfield  {title} {\bibinfo {title} {{Jahn-Teller Effect in Systems with Strong On-Site Spin-Orbit Coupling}},\ }\href {https://doi.org/10.1103/PhysRevLett.116.106401} {\bibfield  {journal} {\bibinfo  {journal} {Phys. Rev. Lett.}\ }\textbf {\bibinfo {volume} {116}},\ \bibinfo {pages} {106401} (\bibinfo {year} {2016})}\BibitemShut {NoStop}%
\bibitem [{\citenamefont {Liu}\ and\ \citenamefont {Khaliullin}(2019)}]{Liu2019}%
  \BibitemOpen
  \bibfield  {author} {\bibinfo {author} {\bibfnamefont {H.}~\bibnamefont {Liu}}\ and\ \bibinfo {author} {\bibfnamefont {G.}~\bibnamefont {Khaliullin}},\ }\bibfield  {title} {\bibinfo {title} {{Pseudo-Jahn-Teller Effect and Magnetoelastic Coupling in Spin-Orbit Mott Insulators}},\ }\href {https://doi.org/10.1103/PhysRevLett.122.057203} {\bibfield  {journal} {\bibinfo  {journal} {Phys. Rev. Lett.}\ }\textbf {\bibinfo {volume} {122}},\ \bibinfo {pages} {057203} (\bibinfo {year} {2019})}\BibitemShut {NoStop}%
\bibitem [{\citenamefont {Streltsov}\ and\ \citenamefont {Khomskii}(2020)}]{Streltsov2020}%
  \BibitemOpen
  \bibfield  {author} {\bibinfo {author} {\bibfnamefont {S.~V.}\ \bibnamefont {Streltsov}}\ and\ \bibinfo {author} {\bibfnamefont {D.~I.}\ \bibnamefont {Khomskii}},\ }\bibfield  {title} {\bibinfo {title} {{Jahn-Teller Effect and Spin-Orbit Coupling: Friends or Foes?}},\ }\href {https://doi.org/10.1103/PhysRevX.10.031043} {\bibfield  {journal} {\bibinfo  {journal} {Phys. Rev. X}\ }\textbf {\bibinfo {volume} {10}},\ \bibinfo {pages} {031043} (\bibinfo {year} {2020})}\BibitemShut {NoStop}%
\bibitem [{\citenamefont {Hirai}\ and\ \citenamefont {Hiroi}(2019)}]{Hirai2019}%
  \BibitemOpen
  \bibfield  {author} {\bibinfo {author} {\bibfnamefont {D.}~\bibnamefont {Hirai}}\ and\ \bibinfo {author} {\bibfnamefont {Z.}~\bibnamefont {Hiroi}},\ }\bibfield  {title} {\bibinfo {title} {{Successive Symmetry Breaking in a Jeff = 3/2 Quartet in the Spin–Orbit Coupled Insulator Ba$_2$MgReO$_6$}},\ }\href {https://doi.org/10.7566/JPSJ.88.064712} {\bibfield  {journal} {\bibinfo  {journal} {Journal of the Physical Society of Japan}\ }\textbf {\bibinfo {volume} {88}},\ \bibinfo {pages} {064712} (\bibinfo {year} {2019})}\BibitemShut {NoStop}%
\bibitem [{\citenamefont {Hirai}\ \emph {et~al.}(2020)\citenamefont {Hirai}, \citenamefont {Sagayama}, \citenamefont {Gao}, \citenamefont {Ohsumi}, \citenamefont {Chen}, \citenamefont {Arima},\ and\ \citenamefont {Hiroi}}]{Hirai2020}%
  \BibitemOpen
  \bibfield  {author} {\bibinfo {author} {\bibfnamefont {D.}~\bibnamefont {Hirai}}, \bibinfo {author} {\bibfnamefont {H.}~\bibnamefont {Sagayama}}, \bibinfo {author} {\bibfnamefont {S.}~\bibnamefont {Gao}}, \bibinfo {author} {\bibfnamefont {H.}~\bibnamefont {Ohsumi}}, \bibinfo {author} {\bibfnamefont {G.}~\bibnamefont {Chen}}, \bibinfo {author} {\bibfnamefont {T.-h.}\ \bibnamefont {Arima}},\ and\ \bibinfo {author} {\bibfnamefont {Z.}~\bibnamefont {Hiroi}},\ }\bibfield  {title} {\bibinfo {title} {{Detection of multipolar orders in the spin-orbit-coupled $5d$ Mott insulator $\mathrm{B}{\mathrm{a}}_{2}\mathrm{MgRe}{\mathrm{O}}_{6}$}},\ }\href {https://doi.org/10.1103/PhysRevResearch.2.022063} {\bibfield  {journal} {\bibinfo  {journal} {Phys. Rev. Res.}\ }\textbf {\bibinfo {volume} {2}},\ \bibinfo {pages} {022063} (\bibinfo {year} {2020})}\BibitemShut {NoStop}%
\bibitem [{\citenamefont {Frontini}\ \emph {et~al.}(2023)\citenamefont {Frontini}, \citenamefont {Johnstone}, \citenamefont {Iwahara}, \citenamefont {Bhattacharyya}, \citenamefont {Bogdanov}, \citenamefont {Hozoi}, \citenamefont {Upton}, \citenamefont {Casa}, \citenamefont {Hirai},\ and\ \citenamefont {Kim}}]{Frontini2023}%
  \BibitemOpen
  \bibfield  {author} {\bibinfo {author} {\bibfnamefont {F.~I.}\ \bibnamefont {Frontini}}, \bibinfo {author} {\bibfnamefont {G.~H.}\ \bibnamefont {Johnstone}}, \bibinfo {author} {\bibfnamefont {N.}~\bibnamefont {Iwahara}}, \bibinfo {author} {\bibfnamefont {P.}~\bibnamefont {Bhattacharyya}}, \bibinfo {author} {\bibfnamefont {N.~A.}\ \bibnamefont {Bogdanov}}, \bibinfo {author} {\bibfnamefont {L.}~\bibnamefont {Hozoi}}, \bibinfo {author} {\bibfnamefont {M.~H.}\ \bibnamefont {Upton}}, \bibinfo {author} {\bibfnamefont {D.~M.}\ \bibnamefont {Casa}}, \bibinfo {author} {\bibfnamefont {D.}~\bibnamefont {Hirai}},\ and\ \bibinfo {author} {\bibfnamefont {Y.-J.}\ \bibnamefont {Kim}},\ }\href@noop {} {\bibinfo {title} {{Spin-orbit-lattice entangled state in A$_2$MgReO$_6$ (A = Ca, Sr, Ba) revealed by resonant inelastic X-ray scattering}}} (\bibinfo {year} {2023}),\ \Eprint {https://arxiv.org/abs/2311.01621} {arXiv:2311.01621 [cond-mat.str-el]} \BibitemShut {NoStop}%
\bibitem [{\citenamefont {Agrestini}\ \emph {et~al.}(2024)\citenamefont {Agrestini}, \citenamefont {Borgatti}, \citenamefont {Florio}, \citenamefont {Frassineti}, \citenamefont {Mosca}, \citenamefont {Faure}, \citenamefont {Detlefs}, \citenamefont {Sahle}, \citenamefont {Francoual}, \citenamefont {Choi}, \citenamefont {Garcia-Fernandez}, \citenamefont {Zhou}, \citenamefont {Mitrovic}, \citenamefont {Woodward}, \citenamefont {Ghiringhelli}, \citenamefont {Franchini}, \citenamefont {Boscherini}, \citenamefont {Sanna}, ,\ and\ \citenamefont {Sala}}]{Agrestini2024}%
  \BibitemOpen
  \bibfield  {author} {\bibinfo {author} {\bibfnamefont {S.}~\bibnamefont {Agrestini}}, \bibinfo {author} {\bibfnamefont {F.}~\bibnamefont {Borgatti}}, \bibinfo {author} {\bibfnamefont {P.}~\bibnamefont {Florio}}, \bibinfo {author} {\bibfnamefont {J.}~\bibnamefont {Frassineti}}, \bibinfo {author} {\bibfnamefont {D.~F.}\ \bibnamefont {Mosca}}, \bibinfo {author} {\bibfnamefont {Q.}~\bibnamefont {Faure}}, \bibinfo {author} {\bibfnamefont {B.}~\bibnamefont {Detlefs}}, \bibinfo {author} {\bibfnamefont {C.~J.}\ \bibnamefont {Sahle}}, \bibinfo {author} {\bibfnamefont {S.}~\bibnamefont {Francoual}}, \bibinfo {author} {\bibfnamefont {J.}~\bibnamefont {Choi}}, \bibinfo {author} {\bibfnamefont {M.}~\bibnamefont {Garcia-Fernandez}}, \bibinfo {author} {\bibfnamefont {K.-J.}\ \bibnamefont {Zhou}}, \bibinfo {author} {\bibfnamefont {V.~F.}\ \bibnamefont {Mitrovic}}, \bibinfo {author} {\bibfnamefont {P.~M.}\ \bibnamefont {Woodward}}, \bibinfo {author} {\bibfnamefont {G.}~\bibnamefont {Ghiringhelli}}, \bibinfo {author}
  {\bibfnamefont {C.}~\bibnamefont {Franchini}}, \bibinfo {author} {\bibfnamefont {F.}~\bibnamefont {Boscherini}}, \bibinfo {author} {\bibfnamefont {S.}~\bibnamefont {Sanna}}, ,\ and\ \bibinfo {author} {\bibfnamefont {M.~M.}\ \bibnamefont {Sala}},\ }\href@noop {} {\bibinfo {title} {{The origin of magnetism in a supposedly nonmagnetic osmium oxide}}} (\bibinfo {year} {2024}),\ \Eprint {https://arxiv.org/abs/2401.12035} {arXiv:2401.12035 [cond-mat.str-el]} \BibitemShut {NoStop}%
\bibitem [{\citenamefont {Chen}\ \emph {et~al.}(2010)\citenamefont {Chen}, \citenamefont {Pereira},\ and\ \citenamefont {Balents}}]{Chen2010}%
  \BibitemOpen
  \bibfield  {author} {\bibinfo {author} {\bibfnamefont {G.}~\bibnamefont {Chen}}, \bibinfo {author} {\bibfnamefont {R.}~\bibnamefont {Pereira}},\ and\ \bibinfo {author} {\bibfnamefont {L.}~\bibnamefont {Balents}},\ }\bibfield  {title} {\bibinfo {title} {{Exotic phases induced by strong spin-orbit coupling in ordered double perovskites}},\ }\href {https://doi.org/10.1103/PhysRevB.82.174440} {\bibfield  {journal} {\bibinfo  {journal} {Phys. Rev. B}\ }\textbf {\bibinfo {volume} {82}},\ \bibinfo {pages} {174440} (\bibinfo {year} {2010})}\BibitemShut {NoStop}%
\bibitem [{\citenamefont {Lu}\ \emph {et~al.}(2017)\citenamefont {Lu}, \citenamefont {Song}, \citenamefont {Liu}, \citenamefont {Reyes}, \citenamefont {Kuhns}, \citenamefont {Lee}, \citenamefont {Fisher},\ and\ \citenamefont {Mitrović}}]{Lu2017}%
  \BibitemOpen
  \bibfield  {author} {\bibinfo {author} {\bibfnamefont {L.}~\bibnamefont {Lu}}, \bibinfo {author} {\bibfnamefont {M.}~\bibnamefont {Song}}, \bibinfo {author} {\bibfnamefont {W.}~\bibnamefont {Liu}}, \bibinfo {author} {\bibfnamefont {A.~P.}\ \bibnamefont {Reyes}}, \bibinfo {author} {\bibfnamefont {P.}~\bibnamefont {Kuhns}}, \bibinfo {author} {\bibfnamefont {H.~O.}\ \bibnamefont {Lee}}, \bibinfo {author} {\bibfnamefont {I.~R.}\ \bibnamefont {Fisher}},\ and\ \bibinfo {author} {\bibfnamefont {V.~F.}\ \bibnamefont {Mitrović}},\ }\bibfield  {title} {\bibinfo {title} {{Magnetism and local symmetry breaking in a Mott insulator with strong spin orbit interactions}},\ }\href {https://doi.org/10.1038/ncomms14407} {\bibfield  {journal} {\bibinfo  {journal} {Nat. Commun.}\ }\textbf {\bibinfo {volume} {8}},\ \bibinfo {pages} {14407} (\bibinfo {year} {2017})}\BibitemShut {NoStop}%
\bibitem [{\citenamefont {Xu}\ \emph {et~al.}(2016)\citenamefont {Xu}, \citenamefont {Bogdanov}, \citenamefont {Princep}, \citenamefont {Fulde}, \citenamefont {van~den Brink},\ and\ \citenamefont {Hozoi}}]{Xu2016}%
  \BibitemOpen
  \bibfield  {author} {\bibinfo {author} {\bibfnamefont {L.}~\bibnamefont {Xu}}, \bibinfo {author} {\bibfnamefont {N.~A.}\ \bibnamefont {Bogdanov}}, \bibinfo {author} {\bibfnamefont {A.}~\bibnamefont {Princep}}, \bibinfo {author} {\bibfnamefont {P.}~\bibnamefont {Fulde}}, \bibinfo {author} {\bibfnamefont {J.}~\bibnamefont {van~den Brink}},\ and\ \bibinfo {author} {\bibfnamefont {L.}~\bibnamefont {Hozoi}},\ }\bibfield  {title} {\bibinfo {title} {{Covalency and vibronic couplings make a nonmagnetic j=3/2 ion magnetic}},\ }\href {https://doi.org/10.1038/npjquantmats.2016.29} {\bibfield  {journal} {\bibinfo  {journal} {npj Quantum Materials}\ }\textbf {\bibinfo {volume} {1}},\ \bibinfo {pages} {16029} (\bibinfo {year} {2016})}\BibitemShut {NoStop}%
\bibitem [{\citenamefont {Svoboda}\ \emph {et~al.}(2021)\citenamefont {Svoboda}, \citenamefont {Zhang}, \citenamefont {Randeria},\ and\ \citenamefont {Trivedi}}]{Svoboda2021}%
  \BibitemOpen
  \bibfield  {author} {\bibinfo {author} {\bibfnamefont {C.}~\bibnamefont {Svoboda}}, \bibinfo {author} {\bibfnamefont {W.}~\bibnamefont {Zhang}}, \bibinfo {author} {\bibfnamefont {M.}~\bibnamefont {Randeria}},\ and\ \bibinfo {author} {\bibfnamefont {N.}~\bibnamefont {Trivedi}},\ }\bibfield  {title} {\bibinfo {title} {{Orbital order drives magnetic order in $5{d}^{1}$ and $5{d}^{2}$ double perovskite Mott insulators}},\ }\href {https://doi.org/10.1103/PhysRevB.104.024437} {\bibfield  {journal} {\bibinfo  {journal} {Phys. Rev. B}\ }\textbf {\bibinfo {volume} {104}},\ \bibinfo {pages} {024437} (\bibinfo {year} {2021})}\BibitemShut {NoStop}%
\bibitem [{\citenamefont {Iwahara}\ and\ \citenamefont {Chibotaru}(2023)}]{Iwahara2023}%
  \BibitemOpen
  \bibfield  {author} {\bibinfo {author} {\bibfnamefont {N.}~\bibnamefont {Iwahara}}\ and\ \bibinfo {author} {\bibfnamefont {L.~F.}\ \bibnamefont {Chibotaru}},\ }\bibfield  {title} {\bibinfo {title} {{Vibronic order and emergent magnetism in cubic ${d}^{1}$ double perovskites}},\ }\href {https://doi.org/10.1103/PhysRevB.107.L220404} {\bibfield  {journal} {\bibinfo  {journal} {Phys. Rev. B}\ }\textbf {\bibinfo {volume} {107}},\ \bibinfo {pages} {L220404} (\bibinfo {year} {2023})}\BibitemShut {NoStop}%
\bibitem [{\citenamefont {Razavy}(2014)}]{Razavy2014}%
  \BibitemOpen
  \bibfield  {author} {\bibinfo {author} {\bibfnamefont {M.}~\bibnamefont {Razavy}},\ }\href@noop {} {\emph {\bibinfo {title} {{Quantum Theory of Tunneling}}}},\ \bibinfo {edition} {2nd}\ ed.\ (\bibinfo  {publisher} {World Scientific,},\ \bibinfo {year} {2014})\BibitemShut {NoStop}%
\bibitem [{\citenamefont {Erickson}\ \emph {et~al.}(2007)\citenamefont {Erickson}, \citenamefont {Misra}, \citenamefont {Miller}, \citenamefont {Gupta}, \citenamefont {Schlesinger}, \citenamefont {Harrison}, \citenamefont {Kim},\ and\ \citenamefont {Fisher}}]{Erickson2007}%
  \BibitemOpen
  \bibfield  {author} {\bibinfo {author} {\bibfnamefont {A.~S.}\ \bibnamefont {Erickson}}, \bibinfo {author} {\bibfnamefont {S.}~\bibnamefont {Misra}}, \bibinfo {author} {\bibfnamefont {G.~J.}\ \bibnamefont {Miller}}, \bibinfo {author} {\bibfnamefont {R.~R.}\ \bibnamefont {Gupta}}, \bibinfo {author} {\bibfnamefont {Z.}~\bibnamefont {Schlesinger}}, \bibinfo {author} {\bibfnamefont {W.~A.}\ \bibnamefont {Harrison}}, \bibinfo {author} {\bibfnamefont {J.~M.}\ \bibnamefont {Kim}},\ and\ \bibinfo {author} {\bibfnamefont {I.~R.}\ \bibnamefont {Fisher}},\ }\bibfield  {title} {\bibinfo {title} {{Ferromagnetism in the Mott Insulator ${\mathrm{Ba}}_{2}{\mathrm{NaOsO}}_{6}$}},\ }\href {https://doi.org/10.1103/PhysRevLett.99.016404} {\bibfield  {journal} {\bibinfo  {journal} {Phys. Rev. Lett.}\ }\textbf {\bibinfo {volume} {99}},\ \bibinfo {pages} {016404} (\bibinfo {year} {2007})}\BibitemShut {NoStop}%
\bibitem [{\citenamefont {Willa}\ \emph {et~al.}(2019)\citenamefont {Willa}, \citenamefont {Willa}, \citenamefont {Welp}, \citenamefont {Fisher}, \citenamefont {Rydh}, \citenamefont {Kwok},\ and\ \citenamefont {Islam}}]{Willa2019}%
  \BibitemOpen
  \bibfield  {author} {\bibinfo {author} {\bibfnamefont {K.}~\bibnamefont {Willa}}, \bibinfo {author} {\bibfnamefont {R.}~\bibnamefont {Willa}}, \bibinfo {author} {\bibfnamefont {U.}~\bibnamefont {Welp}}, \bibinfo {author} {\bibfnamefont {I.~R.}\ \bibnamefont {Fisher}}, \bibinfo {author} {\bibfnamefont {A.}~\bibnamefont {Rydh}}, \bibinfo {author} {\bibfnamefont {W.-K.}\ \bibnamefont {Kwok}},\ and\ \bibinfo {author} {\bibfnamefont {Z.}~\bibnamefont {Islam}},\ }\bibfield  {title} {\bibinfo {title} {{Phase transition preceding magnetic long-range order in the double perovskite ${\mathrm{Ba}}_{2}{\mathrm{NaOsO}}_{6}$}},\ }\href {https://doi.org/10.1103/PhysRevB.100.041108} {\bibfield  {journal} {\bibinfo  {journal} {Phys. Rev. B}\ }\textbf {\bibinfo {volume} {100}},\ \bibinfo {pages} {041108} (\bibinfo {year} {2019})}\BibitemShut {NoStop}%
\bibitem [{\citenamefont {Ishikawa}\ \emph {et~al.}(2019)\citenamefont {Ishikawa}, \citenamefont {Takayama}, \citenamefont {Kremer}, \citenamefont {Nuss}, \citenamefont {Dinnebier}, \citenamefont {Kitagawa}, \citenamefont {Ishii},\ and\ \citenamefont {Takagi}}]{Ishikawa2019}%
  \BibitemOpen
  \bibfield  {author} {\bibinfo {author} {\bibfnamefont {H.}~\bibnamefont {Ishikawa}}, \bibinfo {author} {\bibfnamefont {T.}~\bibnamefont {Takayama}}, \bibinfo {author} {\bibfnamefont {R.~K.}\ \bibnamefont {Kremer}}, \bibinfo {author} {\bibfnamefont {J.}~\bibnamefont {Nuss}}, \bibinfo {author} {\bibfnamefont {R.}~\bibnamefont {Dinnebier}}, \bibinfo {author} {\bibfnamefont {K.}~\bibnamefont {Kitagawa}}, \bibinfo {author} {\bibfnamefont {K.}~\bibnamefont {Ishii}},\ and\ \bibinfo {author} {\bibfnamefont {H.}~\bibnamefont {Takagi}},\ }\bibfield  {title} {\bibinfo {title} {{Ordering of hidden multipoles in spin-orbit entangled $5{d}^{1}$ Ta chlorides}},\ }\href {https://doi.org/10.1103/PhysRevB.100.045142} {\bibfield  {journal} {\bibinfo  {journal} {Phys. Rev. B}\ }\textbf {\bibinfo {volume} {100}},\ \bibinfo {pages} {045142} (\bibinfo {year} {2019})}\BibitemShut {NoStop}%
\bibitem [{\citenamefont {Mansouri~Tehrani}\ \emph {et~al.}(2023)\citenamefont {Mansouri~Tehrani}, \citenamefont {Soh}, \citenamefont {P\'asztorov\'a}, \citenamefont {Merkel}, \citenamefont {\ifmmode \check{Z}\else \v{Z}\fi{}ivkovi\ifmmode~\acute{c}\else \'{c}\fi{}}, \citenamefont {R\o{}nnow},\ and\ \citenamefont {Spaldin}}]{Mansouri2023}%
  \BibitemOpen
  \bibfield  {author} {\bibinfo {author} {\bibfnamefont {A.}~\bibnamefont {Mansouri~Tehrani}}, \bibinfo {author} {\bibfnamefont {J.-R.}\ \bibnamefont {Soh}}, \bibinfo {author} {\bibfnamefont {J.}~\bibnamefont {P\'asztorov\'a}}, \bibinfo {author} {\bibfnamefont {M.~E.}\ \bibnamefont {Merkel}}, \bibinfo {author} {\bibfnamefont {I.}~\bibnamefont {\ifmmode \check{Z}\else \v{Z}\fi{}ivkovi\ifmmode~\acute{c}\else \'{c}\fi{}}}, \bibinfo {author} {\bibfnamefont {H.~M.}\ \bibnamefont {R\o{}nnow}},\ and\ \bibinfo {author} {\bibfnamefont {N.~A.}\ \bibnamefont {Spaldin}},\ }\bibfield  {title} {\bibinfo {title} {Charge multipole correlations and order in ${\mathrm{cs}}_{2}\mathrm{Ta}{\mathrm{cl}}_{6}$},\ }\href {https://doi.org/10.1103/PhysRevResearch.5.L012010} {\bibfield  {journal} {\bibinfo  {journal} {Phys. Rev. Res.}\ }\textbf {\bibinfo {volume} {5}},\ \bibinfo {pages} {L012010} (\bibinfo {year} {2023})}\BibitemShut {NoStop}%
\bibitem [{\citenamefont {Maharaj}\ \emph {et~al.}(2020)\citenamefont {Maharaj}, \citenamefont {Sala}, \citenamefont {Stone}, \citenamefont {Kermarrec}, \citenamefont {Ritter}, \citenamefont {Fauth}, \citenamefont {Marjerrison}, \citenamefont {Greedan}, \citenamefont {Paramekanti},\ and\ \citenamefont {Gaulin}}]{Maharaj2020}%
  \BibitemOpen
  \bibfield  {author} {\bibinfo {author} {\bibfnamefont {D.~D.}\ \bibnamefont {Maharaj}}, \bibinfo {author} {\bibfnamefont {G.}~\bibnamefont {Sala}}, \bibinfo {author} {\bibfnamefont {M.~B.}\ \bibnamefont {Stone}}, \bibinfo {author} {\bibfnamefont {E.}~\bibnamefont {Kermarrec}}, \bibinfo {author} {\bibfnamefont {C.}~\bibnamefont {Ritter}}, \bibinfo {author} {\bibfnamefont {F.}~\bibnamefont {Fauth}}, \bibinfo {author} {\bibfnamefont {C.~A.}\ \bibnamefont {Marjerrison}}, \bibinfo {author} {\bibfnamefont {J.~E.}\ \bibnamefont {Greedan}}, \bibinfo {author} {\bibfnamefont {A.}~\bibnamefont {Paramekanti}},\ and\ \bibinfo {author} {\bibfnamefont {B.~D.}\ \bibnamefont {Gaulin}},\ }\bibfield  {title} {\bibinfo {title} {{Octupolar versus N\'eel Order in Cubic $5{d}^{2}$ Double Perovskites}},\ }\href {https://doi.org/10.1103/PhysRevLett.124.087206} {\bibfield  {journal} {\bibinfo  {journal} {Phys. Rev. Lett.}\ }\textbf {\bibinfo {volume} {124}},\ \bibinfo {pages} {087206} (\bibinfo {year} {2020})}\BibitemShut {NoStop}%
\bibitem [{\citenamefont {Kermarrec}\ \emph {et~al.}(2015)\citenamefont {Kermarrec}, \citenamefont {Marjerrison}, \citenamefont {Thompson}, \citenamefont {Maharaj}, \citenamefont {Levin}, \citenamefont {Kroeker}, \citenamefont {Granroth}, \citenamefont {Flacau}, \citenamefont {Yamani}, \citenamefont {Greedan},\ and\ \citenamefont {Gaulin}}]{Kermarrec2015}%
  \BibitemOpen
  \bibfield  {author} {\bibinfo {author} {\bibfnamefont {E.}~\bibnamefont {Kermarrec}}, \bibinfo {author} {\bibfnamefont {C.~A.}\ \bibnamefont {Marjerrison}}, \bibinfo {author} {\bibfnamefont {C.~M.}\ \bibnamefont {Thompson}}, \bibinfo {author} {\bibfnamefont {D.~D.}\ \bibnamefont {Maharaj}}, \bibinfo {author} {\bibfnamefont {K.}~\bibnamefont {Levin}}, \bibinfo {author} {\bibfnamefont {S.}~\bibnamefont {Kroeker}}, \bibinfo {author} {\bibfnamefont {G.~E.}\ \bibnamefont {Granroth}}, \bibinfo {author} {\bibfnamefont {R.}~\bibnamefont {Flacau}}, \bibinfo {author} {\bibfnamefont {Z.}~\bibnamefont {Yamani}}, \bibinfo {author} {\bibfnamefont {J.~E.}\ \bibnamefont {Greedan}},\ and\ \bibinfo {author} {\bibfnamefont {B.~D.}\ \bibnamefont {Gaulin}},\ }\bibfield  {title} {\bibinfo {title} {{Frustrated fcc antiferromagnet ${\mathrm{Ba}}_{2}{\mathrm{YOsO}}_{6}$: Structural characterization, magnetic properties, and neutron scattering studies}},\ }\href {https://doi.org/10.1103/PhysRevB.91.075133} {\bibfield  {journal}
  {\bibinfo  {journal} {Phys. Rev. B}\ }\textbf {\bibinfo {volume} {91}},\ \bibinfo {pages} {075133} (\bibinfo {year} {2015})}\BibitemShut {NoStop}%
\bibitem [{\citenamefont {Marjerrison}\ \emph {et~al.}(2016)\citenamefont {Marjerrison}, \citenamefont {Thompson}, \citenamefont {Sala}, \citenamefont {Maharaj}, \citenamefont {Kermarrec}, \citenamefont {Cai}, \citenamefont {Hallas}, \citenamefont {Wilson}, \citenamefont {Munsie}, \citenamefont {Granroth}, \citenamefont {Flacau}, \citenamefont {Greedan}, \citenamefont {Gaulin},\ and\ \citenamefont {Luke}}]{Marjerrison2016}%
  \BibitemOpen
  \bibfield  {author} {\bibinfo {author} {\bibfnamefont {C.~A.}\ \bibnamefont {Marjerrison}}, \bibinfo {author} {\bibfnamefont {C.~M.}\ \bibnamefont {Thompson}}, \bibinfo {author} {\bibfnamefont {G.}~\bibnamefont {Sala}}, \bibinfo {author} {\bibfnamefont {D.~D.}\ \bibnamefont {Maharaj}}, \bibinfo {author} {\bibfnamefont {E.}~\bibnamefont {Kermarrec}}, \bibinfo {author} {\bibfnamefont {Y.}~\bibnamefont {Cai}}, \bibinfo {author} {\bibfnamefont {A.~M.}\ \bibnamefont {Hallas}}, \bibinfo {author} {\bibfnamefont {M.~N.}\ \bibnamefont {Wilson}}, \bibinfo {author} {\bibfnamefont {T.~J.~S.}\ \bibnamefont {Munsie}}, \bibinfo {author} {\bibfnamefont {G.~E.}\ \bibnamefont {Granroth}}, \bibinfo {author} {\bibfnamefont {R.}~\bibnamefont {Flacau}}, \bibinfo {author} {\bibfnamefont {J.~E.}\ \bibnamefont {Greedan}}, \bibinfo {author} {\bibfnamefont {B.~D.}\ \bibnamefont {Gaulin}},\ and\ \bibinfo {author} {\bibfnamefont {G.~M.}\ \bibnamefont {Luke}},\ }\bibfield  {title} {\bibinfo {title} {{Cubic Re6+ (5d1) Double Perovskites,
  Ba2MgReO6, Ba2ZnReO6, and Ba2Y2/3ReO6: Magnetism, Heat Capacity, uSR, and Neutron Scattering Studies and Comparison with Theory}},\ }\href {https://doi.org/10.1021/acs.inorgchem.6b01933} {\bibfield  {journal} {\bibinfo  {journal} {Inorganic Chemistry}\ }\textbf {\bibinfo {volume} {55}},\ \bibinfo {pages} {10701} (\bibinfo {year} {2016})}\BibitemShut {NoStop}%
\bibitem [{\citenamefont {Ishikawa}\ \emph {et~al.}(2021)\citenamefont {Ishikawa}, \citenamefont {Hirai}, \citenamefont {Ikeda}, \citenamefont {Gen}, \citenamefont {Yajima}, \citenamefont {Matsuo}, \citenamefont {Matsuda}, \citenamefont {Hiroi},\ and\ \citenamefont {Kindo}}]{Ishikawa2021}%
  \BibitemOpen
  \bibfield  {author} {\bibinfo {author} {\bibfnamefont {H.}~\bibnamefont {Ishikawa}}, \bibinfo {author} {\bibfnamefont {D.}~\bibnamefont {Hirai}}, \bibinfo {author} {\bibfnamefont {A.}~\bibnamefont {Ikeda}}, \bibinfo {author} {\bibfnamefont {M.}~\bibnamefont {Gen}}, \bibinfo {author} {\bibfnamefont {T.}~\bibnamefont {Yajima}}, \bibinfo {author} {\bibfnamefont {A.}~\bibnamefont {Matsuo}}, \bibinfo {author} {\bibfnamefont {Y.~H.}\ \bibnamefont {Matsuda}}, \bibinfo {author} {\bibfnamefont {Z.}~\bibnamefont {Hiroi}},\ and\ \bibinfo {author} {\bibfnamefont {K.}~\bibnamefont {Kindo}},\ }\bibfield  {title} {\bibinfo {title} {{Phase transition in the $5{d}^{1}$ double perovskite ${\mathrm{Ba}}_{2}{\mathrm{CaReO}}_{6}$ induced by high magnetic field}},\ }\href {https://doi.org/10.1103/PhysRevB.104.174422} {\bibfield  {journal} {\bibinfo  {journal} {Phys. Rev. B}\ }\textbf {\bibinfo {volume} {104}},\ \bibinfo {pages} {174422} (\bibinfo {year} {2021})}\BibitemShut {NoStop}%
\bibitem [{\citenamefont {Pásztorová}\ \emph {et~al.}(2023)\citenamefont {Pásztorová}, \citenamefont {Tehrani}, \citenamefont {Živković}, \citenamefont {Spaldin},\ and\ \citenamefont {Rønnow}}]{Pasztorova2023}%
  \BibitemOpen
  \bibfield  {author} {\bibinfo {author} {\bibfnamefont {J.}~\bibnamefont {Pásztorová}}, \bibinfo {author} {\bibfnamefont {A.~M.}\ \bibnamefont {Tehrani}}, \bibinfo {author} {\bibfnamefont {I.}~\bibnamefont {Živković}}, \bibinfo {author} {\bibfnamefont {N.~A.}\ \bibnamefont {Spaldin}},\ and\ \bibinfo {author} {\bibfnamefont {H.~M.}\ \bibnamefont {Rønnow}},\ }\bibfield  {title} {\bibinfo {title} {{Experimental and theoretical thermodynamic studies in Ba2MgReO6—the ground state in the context of Jahn-Teller effect}},\ }\href {https://doi.org/10.1088/1361-648X/acc62a} {\bibfield  {journal} {\bibinfo  {journal} {Journal of Physics: Condensed Matter}\ }\textbf {\bibinfo {volume} {35}},\ \bibinfo {pages} {245603} (\bibinfo {year} {2023})}\BibitemShut {NoStop}%
\bibitem [{\citenamefont {{Peysson, Y.}}\ \emph {et~al.}(1986)\citenamefont {{Peysson, Y.}}, \citenamefont {{Ayache, C.}}, \citenamefont {{Rossat-Mignod, J.}}, \citenamefont {{Kunii, S.}},\ and\ \citenamefont {{Kasuya, T.}}}]{Peysson1986}%
  \BibitemOpen
  \bibfield  {author} {\bibinfo {author} {\bibnamefont {{Peysson, Y.}}}, \bibinfo {author} {\bibnamefont {{Ayache, C.}}}, \bibinfo {author} {\bibnamefont {{Rossat-Mignod, J.}}}, \bibinfo {author} {\bibnamefont {{Kunii, S.}}},\ and\ \bibinfo {author} {\bibnamefont {{Kasuya, T.}}},\ }\bibfield  {title} {\bibinfo {title} {{High magnetic field study of the specific heat of CeB$_6$ and LaB$_6$}},\ }\href {https://doi.org/10.1051/jphys:01986004701011300} {\bibfield  {journal} {\bibinfo  {journal} {J. Phys. France}\ }\textbf {\bibinfo {volume} {47}},\ \bibinfo {pages} {113} (\bibinfo {year} {1986})}\BibitemShut {NoStop}%
\bibitem [{\citenamefont {Amara}(2019)}]{Amara2020}%
  \BibitemOpen
  \bibfield  {author} {\bibinfo {author} {\bibfnamefont {M.}~\bibnamefont {Amara}},\ }\bibfield  {title} {\bibinfo {title} {{Dynamical splitting of cubic crystal field levels in rare-earth cage compounds}},\ }\href {https://doi.org/10.1103/PhysRevB.99.174405} {\bibfield  {journal} {\bibinfo  {journal} {Phys. Rev. B}\ }\textbf {\bibinfo {volume} {99}},\ \bibinfo {pages} {174405} (\bibinfo {year} {2019})}\BibitemShut {NoStop}%
\bibitem [{\citenamefont {Shiina}\ \emph {et~al.}(1997)\citenamefont {Shiina}, \citenamefont {Shiba},\ and\ \citenamefont {Thalmeier}}]{Shiina1997}%
  \BibitemOpen
  \bibfield  {author} {\bibinfo {author} {\bibfnamefont {R.}~\bibnamefont {Shiina}}, \bibinfo {author} {\bibfnamefont {H.}~\bibnamefont {Shiba}},\ and\ \bibinfo {author} {\bibfnamefont {P.}~\bibnamefont {Thalmeier}},\ }\bibfield  {title} {\bibinfo {title} {{Magnetic-Field Effects on Quadrupolar Ordering in a $\Gamma_8$-Quartet System CeB$_6$}},\ }\href {https://doi.org/10.1143/JPSJ.66.1741} {\bibfield  {journal} {\bibinfo  {journal} {Journal of the Physical Society of Japan}\ }\textbf {\bibinfo {volume} {66}},\ \bibinfo {pages} {1741} (\bibinfo {year} {1997})}\BibitemShut {NoStop}%
\bibitem [{\citenamefont {Matsumura}\ \emph {et~al.}(2009)\citenamefont {Matsumura}, \citenamefont {Yonemura}, \citenamefont {Kunimori}, \citenamefont {Sera},\ and\ \citenamefont {Iga}}]{Matsumara2009}%
  \BibitemOpen
  \bibfield  {author} {\bibinfo {author} {\bibfnamefont {T.}~\bibnamefont {Matsumura}}, \bibinfo {author} {\bibfnamefont {T.}~\bibnamefont {Yonemura}}, \bibinfo {author} {\bibfnamefont {K.}~\bibnamefont {Kunimori}}, \bibinfo {author} {\bibfnamefont {M.}~\bibnamefont {Sera}},\ and\ \bibinfo {author} {\bibfnamefont {F.}~\bibnamefont {Iga}},\ }\bibfield  {title} {\bibinfo {title} {{Magnetic Field Induced $4f$ Octupole in ${\mathrm{CeB}}_{6}$ Probed by Resonant X-Ray Diffraction}},\ }\href {https://doi.org/10.1103/PhysRevLett.103.017203} {\bibfield  {journal} {\bibinfo  {journal} {Phys. Rev. Lett.}\ }\textbf {\bibinfo {volume} {103}},\ \bibinfo {pages} {017203} (\bibinfo {year} {2009})}\BibitemShut {NoStop}%
\bibitem [{\citenamefont {Soh}\ \emph {et~al.}(2023)\citenamefont {Soh}, \citenamefont {Merkel}, \citenamefont {Pourovskii}, \citenamefont {Živković}, \citenamefont {Malanyuk}, \citenamefont {Pásztorová}, \citenamefont {Francoual}, \citenamefont {Hirai}, \citenamefont {Urru}, \citenamefont {Tolj}, \citenamefont {Fiore-Mosca}, \citenamefont {Yazyev}, \citenamefont {Spaldin}, \citenamefont {Ederer},\ and\ \citenamefont {Rønnow}}]{Soh2023}%
  \BibitemOpen
  \bibfield  {author} {\bibinfo {author} {\bibfnamefont {J.-R.}\ \bibnamefont {Soh}}, \bibinfo {author} {\bibfnamefont {M.~E.}\ \bibnamefont {Merkel}}, \bibinfo {author} {\bibfnamefont {L.}~\bibnamefont {Pourovskii}}, \bibinfo {author} {\bibfnamefont {I.}~\bibnamefont {Živković}}, \bibinfo {author} {\bibfnamefont {O.}~\bibnamefont {Malanyuk}}, \bibinfo {author} {\bibfnamefont {J.}~\bibnamefont {Pásztorová}}, \bibinfo {author} {\bibfnamefont {S.}~\bibnamefont {Francoual}}, \bibinfo {author} {\bibfnamefont {D.}~\bibnamefont {Hirai}}, \bibinfo {author} {\bibfnamefont {A.}~\bibnamefont {Urru}}, \bibinfo {author} {\bibfnamefont {D.}~\bibnamefont {Tolj}}, \bibinfo {author} {\bibfnamefont {D.}~\bibnamefont {Fiore-Mosca}}, \bibinfo {author} {\bibfnamefont {O.}~\bibnamefont {Yazyev}}, \bibinfo {author} {\bibfnamefont {N.~A.}\ \bibnamefont {Spaldin}}, \bibinfo {author} {\bibfnamefont {C.}~\bibnamefont {Ederer}},\ and\ \bibinfo {author} {\bibfnamefont {H.~M.}\ \bibnamefont {Rønnow}},\ }\href@noop {} {\bibinfo
  {title} {{Spectroscopic signatures and origin of a hidden order in Ba$_2$MgReO$_6$}}} (\bibinfo {year} {2023}),\ \Eprint {https://arxiv.org/abs/2312.01767} {arXiv:2312.01767 [cond-mat.str-el]} \BibitemShut {NoStop}%
\bibitem [{\citenamefont {Mansouri~Tehrani}\ and\ \citenamefont {Spaldin}(2021)}]{Mansouri2021}%
  \BibitemOpen
  \bibfield  {author} {\bibinfo {author} {\bibfnamefont {A.}~\bibnamefont {Mansouri~Tehrani}}\ and\ \bibinfo {author} {\bibfnamefont {N.~A.}\ \bibnamefont {Spaldin}},\ }\bibfield  {title} {\bibinfo {title} {{Untangling the structural, magnetic dipole, and charge multipolar orders in ${\mathrm{Ba}}_{2}{\mathrm{MgReO}}_{6}$}},\ }\href {https://doi.org/10.1103/PhysRevMaterials.5.104410} {\bibfield  {journal} {\bibinfo  {journal} {Phys. Rev. Mater.}\ }\textbf {\bibinfo {volume} {5}},\ \bibinfo {pages} {104410} (\bibinfo {year} {2021})}\BibitemShut {NoStop}%
\bibitem [{\citenamefont {Iwanaga}\ \emph {et~al.}(1999)\citenamefont {Iwanaga}, \citenamefont {Inaguma},\ and\ \citenamefont {Itoh}}]{Iwanaga1999}%
  \BibitemOpen
  \bibfield  {author} {\bibinfo {author} {\bibfnamefont {D.}~\bibnamefont {Iwanaga}}, \bibinfo {author} {\bibfnamefont {Y.}~\bibnamefont {Inaguma}},\ and\ \bibinfo {author} {\bibfnamefont {M.}~\bibnamefont {Itoh}},\ }\bibfield  {title} {\bibinfo {title} {Crystal structure and magnetic properties of b-site ordered perovskite-type oxides a2cubo6 (a=ba, sr; b=w, te)},\ }\href {https://doi.org/10.1006/jssc.1999.8273} {\bibfield  {journal} {\bibinfo  {journal} {Journal of Solid State Chemistry}\ }\textbf {\bibinfo {volume} {147}},\ \bibinfo {pages} {291} (\bibinfo {year} {1999})}\BibitemShut {NoStop}%
\bibitem [{\citenamefont {Iwahara}\ \emph {et~al.}(2018)\citenamefont {Iwahara}, \citenamefont {Vieru},\ and\ \citenamefont {Chibotaru}}]{Iwahara2018}%
  \BibitemOpen
  \bibfield  {author} {\bibinfo {author} {\bibfnamefont {N.}~\bibnamefont {Iwahara}}, \bibinfo {author} {\bibfnamefont {V.}~\bibnamefont {Vieru}},\ and\ \bibinfo {author} {\bibfnamefont {L.~F.}\ \bibnamefont {Chibotaru}},\ }\bibfield  {title} {\bibinfo {title} {{Spin-orbital-lattice entangled states in cubic ${d}^{1}$ double perovskites}},\ }\href {https://doi.org/10.1103/PhysRevB.98.075138} {\bibfield  {journal} {\bibinfo  {journal} {Phys. Rev. B}\ }\textbf {\bibinfo {volume} {98}},\ \bibinfo {pages} {075138} (\bibinfo {year} {2018})}\BibitemShut {NoStop}%
\bibitem [{\citenamefont {Lovesey}\ and\ \citenamefont {Khalyavin}(2021)}]{Lovesey2021}%
  \BibitemOpen
  \bibfield  {author} {\bibinfo {author} {\bibfnamefont {S.~W.}\ \bibnamefont {Lovesey}}\ and\ \bibinfo {author} {\bibfnamefont {D.~D.}\ \bibnamefont {Khalyavin}},\ }\bibfield  {title} {\bibinfo {title} {Magnetic order and $5{d}^{1}$ multipoles in a rhenate double perovskite ${\mathrm{ba}}_{2}{\mathrm{mgreo}}_{6}$},\ }\href {https://doi.org/10.1103/PhysRevB.103.235160} {\bibfield  {journal} {\bibinfo  {journal} {Phys. Rev. B}\ }\textbf {\bibinfo {volume} {103}},\ \bibinfo {pages} {235160} (\bibinfo {year} {2021})}\BibitemShut {NoStop}%
\bibitem [{\citenamefont {Strocov}\ \emph {et~al.}(2010)\citenamefont {Strocov}, \citenamefont {Schmitt}, \citenamefont {Flechsig}, \citenamefont {Schmidt}, \citenamefont {Imhof}, \citenamefont {Chen}, \citenamefont {Raabe}, \citenamefont {Betemps}, \citenamefont {Zimoch}, \citenamefont {Krempasky} \emph {et~al.}}]{Strocov2010}%
  \BibitemOpen
  \bibfield  {author} {\bibinfo {author} {\bibfnamefont {V.}~\bibnamefont {Strocov}}, \bibinfo {author} {\bibfnamefont {T.}~\bibnamefont {Schmitt}}, \bibinfo {author} {\bibfnamefont {U.}~\bibnamefont {Flechsig}}, \bibinfo {author} {\bibfnamefont {T.}~\bibnamefont {Schmidt}}, \bibinfo {author} {\bibfnamefont {A.}~\bibnamefont {Imhof}}, \bibinfo {author} {\bibfnamefont {Q.}~\bibnamefont {Chen}}, \bibinfo {author} {\bibfnamefont {J.}~\bibnamefont {Raabe}}, \bibinfo {author} {\bibfnamefont {R.}~\bibnamefont {Betemps}}, \bibinfo {author} {\bibfnamefont {D.}~\bibnamefont {Zimoch}}, \bibinfo {author} {\bibfnamefont {J.}~\bibnamefont {Krempasky}}, \emph {et~al.},\ }\bibfield  {title} {\bibinfo {title} {High-resolution soft x-ray beamline adress at the swiss light source for resonant inelastic x-ray scattering and angle-resolved photoelectron spectroscopies},\ }\href@noop {} {\bibfield  {journal} {\bibinfo  {journal} {Journal of synchrotron radiation}\ }\textbf {\bibinfo {volume} {17}},\ \bibinfo {pages} {631}
  (\bibinfo {year} {2010})}\BibitemShut {NoStop}%
\bibitem [{\citenamefont {Werner}\ \emph {et~al.}(2012)\citenamefont {Werner}, \citenamefont {Knowles}, \citenamefont {Knizia}, \citenamefont {Manby},\ and\ \citenamefont {Schuetz}}]{Molpro}%
  \BibitemOpen
  \bibfield  {author} {\bibinfo {author} {\bibfnamefont {H.-J.}\ \bibnamefont {Werner}}, \bibinfo {author} {\bibfnamefont {P.~J.}\ \bibnamefont {Knowles}}, \bibinfo {author} {\bibfnamefont {G.}~\bibnamefont {Knizia}}, \bibinfo {author} {\bibfnamefont {F.~R.}\ \bibnamefont {Manby}},\ and\ \bibinfo {author} {\bibfnamefont {M.}~\bibnamefont {Schuetz}},\ }\bibfield  {title} {\bibinfo {title} {Molpro: a general-purpose quantum chemistry program package},\ }\href {https://doi.org/10.1002/wcms.82} {\bibfield  {journal} {\bibinfo  {journal} {Wiley Interdisciplinary Reviews: Computational Molecular Science}\ }\textbf {\bibinfo {volume} {2}},\ \bibinfo {pages} {242} (\bibinfo {year} {2012})}\BibitemShut {NoStop}%
\bibitem [{\citenamefont {Figgen}\ \emph {et~al.}(2009)\citenamefont {Figgen}, \citenamefont {Peterson}, \citenamefont {Dolg},\ and\ \citenamefont {Stoll}}]{ReBasis}%
  \BibitemOpen
  \bibfield  {author} {\bibinfo {author} {\bibfnamefont {D.}~\bibnamefont {Figgen}}, \bibinfo {author} {\bibfnamefont {K.~A.}\ \bibnamefont {Peterson}}, \bibinfo {author} {\bibfnamefont {M.}~\bibnamefont {Dolg}},\ and\ \bibinfo {author} {\bibfnamefont {H.}~\bibnamefont {Stoll}},\ }\bibfield  {title} {\bibinfo {title} {{Energy-consistent pseudopotentials and correlation consistent basis sets for the 5d elements {Hf-Pt}}},\ }\href@noop {} {\bibfield  {journal} {\bibinfo  {journal} {J Chem Phys}\ }\textbf {\bibinfo {volume} {130}},\ \bibinfo {pages} {164108} (\bibinfo {year} {2009})}\BibitemShut {NoStop}%
\bibitem [{\citenamefont {Peterson}\ and\ \citenamefont {Dunning}(2002)}]{OBasis}%
  \BibitemOpen
  \bibfield  {author} {\bibinfo {author} {\bibfnamefont {K.~A.}\ \bibnamefont {Peterson}}\ and\ \bibinfo {author} {\bibfnamefont {J.}~\bibnamefont {Dunning}, \bibfnamefont {Thom~H.}},\ }\bibfield  {title} {\bibinfo {title} {{Accurate correlation consistent basis sets for molecular core–valence correlation effects: The second row atoms Al–Ar, and the first row atoms B–Ne revisited}},\ }\href {https://doi.org/10.1063/1.1520138} {\bibfield  {journal} {\bibinfo  {journal} {The Journal of Chemical Physics}\ }\textbf {\bibinfo {volume} {117}},\ \bibinfo {pages} {10548} (\bibinfo {year} {2002})}\BibitemShut {NoStop}%
\bibitem [{\citenamefont {Fuentealba}\ \emph {et~al.}(1985)\citenamefont {Fuentealba}, \citenamefont {von Szentpaly}, \citenamefont {Preuss},\ and\ \citenamefont {Stoll}}]{MgBasis1}%
  \BibitemOpen
  \bibfield  {author} {\bibinfo {author} {\bibfnamefont {P.}~\bibnamefont {Fuentealba}}, \bibinfo {author} {\bibfnamefont {L.}~\bibnamefont {von Szentpaly}}, \bibinfo {author} {\bibfnamefont {H.}~\bibnamefont {Preuss}},\ and\ \bibinfo {author} {\bibfnamefont {H.}~\bibnamefont {Stoll}},\ }\bibfield  {title} {\bibinfo {title} {Pseudopotential calculations for alkaline-earth atoms},\ }\href {https://doi.org/10.1088/0022-3700/18/7/010} {\bibfield  {journal} {\bibinfo  {journal} {Journal of Physics B: Atomic and Molecular Physics}\ }\textbf {\bibinfo {volume} {18}},\ \bibinfo {pages} {1287} (\bibinfo {year} {1985})}\BibitemShut {NoStop}%
\bibitem [{\citenamefont {Fuentealba}(1998)}]{MgBasis2}%
  \BibitemOpen
  \bibfield  {author} {\bibinfo {author} {\bibfnamefont {P.}~\bibnamefont {Fuentealba}},\ }\href@noop {} {\bibfield  {journal} {\bibinfo  {journal} {Unpublished}\ } (\bibinfo {year} {1998})}\BibitemShut {NoStop}%
\bibitem [{\citenamefont {Klintenberg}\ \emph {et~al.}(2000)\citenamefont {Klintenberg}, \citenamefont {Derenzo},\ and\ \citenamefont {Weber}}]{Ewald}%
  \BibitemOpen
  \bibfield  {author} {\bibinfo {author} {\bibfnamefont {M.}~\bibnamefont {Klintenberg}}, \bibinfo {author} {\bibfnamefont {S.}~\bibnamefont {Derenzo}},\ and\ \bibinfo {author} {\bibfnamefont {M.}~\bibnamefont {Weber}},\ }\bibfield  {title} {\bibinfo {title} {{Accurate crystal fields for embedded cluster calculations}},\ }\href {https://doi.org/https://doi.org/10.1016/S0010-4655(00)00071-0} {\bibfield  {journal} {\bibinfo  {journal} {Computer Physics Communications}\ }\textbf {\bibinfo {volume} {131}},\ \bibinfo {pages} {120} (\bibinfo {year} {2000})}\BibitemShut {NoStop}%
\bibitem [{\citenamefont {Knowles}\ and\ \citenamefont {Werner}(1992)}]{MRCI1}%
  \BibitemOpen
  \bibfield  {author} {\bibinfo {author} {\bibfnamefont {P.~J.}\ \bibnamefont {Knowles}}\ and\ \bibinfo {author} {\bibfnamefont {H.-J.}\ \bibnamefont {Werner}},\ }\bibfield  {title} {\bibinfo {title} {{Internally contracted multiconfiguration-reference configuration interaction calculations for excited states}},\ }\href {https://doi.org/10.1007/BF01117405} {\bibfield  {journal} {\bibinfo  {journal} {Theoretica chimica acta}\ }\textbf {\bibinfo {volume} {84}},\ \bibinfo {pages} {95} (\bibinfo {year} {1992})}\BibitemShut {NoStop}%
\bibitem [{\citenamefont {Werner}\ and\ \citenamefont {Knowles}(1988)}]{MRCI2}%
  \BibitemOpen
  \bibfield  {author} {\bibinfo {author} {\bibfnamefont {H.}~\bibnamefont {Werner}}\ and\ \bibinfo {author} {\bibfnamefont {P.~J.}\ \bibnamefont {Knowles}},\ }\bibfield  {title} {\bibinfo {title} {{An efficient internally contracted multiconfiguration–reference configuration interaction method}},\ }\href {https://doi.org/10.1063/1.455556} {\bibfield  {journal} {\bibinfo  {journal} {The Journal of Chemical Physics}\ }\textbf {\bibinfo {volume} {89}},\ \bibinfo {pages} {5803} (\bibinfo {year} {1988})}\BibitemShut {NoStop}%
\bibitem [{\citenamefont {Berning}\ \emph {et~al.}(2000)\citenamefont {Berning}, \citenamefont {Schweizer}, \citenamefont {Werner}, \citenamefont {Knowles},\ and\ \citenamefont {Palmieri}}]{SOC}%
  \BibitemOpen
  \bibfield  {author} {\bibinfo {author} {\bibfnamefont {A.}~\bibnamefont {Berning}}, \bibinfo {author} {\bibfnamefont {M.}~\bibnamefont {Schweizer}}, \bibinfo {author} {\bibfnamefont {H.-J.}\ \bibnamefont {Werner}}, \bibinfo {author} {\bibfnamefont {P.~J.}\ \bibnamefont {Knowles}},\ and\ \bibinfo {author} {\bibfnamefont {P.}~\bibnamefont {Palmieri}},\ }\bibfield  {title} {\bibinfo {title} {Spin-orbit matrix elements for internally contracted multireference configuration interaction wavefunctions},\ }\href {https://doi.org/10.1080/00268970009483386} {\bibfield  {journal} {\bibinfo  {journal} {Molecular Physics}\ }\textbf {\bibinfo {volume} {98}},\ \bibinfo {pages} {1823} (\bibinfo {year} {2000})}\BibitemShut {NoStop}%
\end{thebibliography}%

\end{document}